\documentclass[sn-nature]{sn-jnl}
\usepackage{graphicx}%
\usepackage{multirow}%
\usepackage{amsmath,amssymb,amsfonts}%
\usepackage{amsthm}%
\usepackage{mathrsfs}%
\usepackage[title]{appendix}%
\usepackage{xcolor}%
\usepackage{textcomp}%
\usepackage{manyfoot}%
{\tiny }\usepackage{booktabs}%
\usepackage{algorithm}%
\usepackage{algorithmicx}%
\usepackage{algpseudocode}%
\usepackage{listings}%
\usepackage{soul}
\usepackage{gensymb}
\usepackage{dcolumn}
\usepackage{xfrac}
\usepackage{bm}
\newcolumntype{d}[1]{D{.}{.}{#1}}

\makeatletter

\begin{document}
	\title{New insight into tuning magnetic phases of $R$Mn$_6$Sn$_6$ kagome metals}
	
	\author[1,2]{\fnm{S.~X.~M.}~\sur{Riberolles}}
	
	\author[1,2]{\fnm{Tianxiong}~\sur{Han}}
	
	\author[1,2]{\fnm{Tyler~J.}~\sur{Slade}}
	
	\author[1,2]{\fnm{J.~M.}~\sur{Wilde}}
	
	\author[1,2]{\fnm{A.}~\sur{Sapkota}}
	
	\author[3]{\fnm{Wei}~\sur{Tian}}
	
	\author[3]{\fnm{Qiang}~\sur{Zhang}}
	
	\author[3]{\fnm{D.~L.}~\sur{Abernathy}}
	
	\author[4,5]{\fnm{L.~D.}~\sur{Sanjeewa}}
	
	\author[1,2]{\fnm{S.~L.}~\sur{Bud'ko}}
	
	\author[1,2]{\fnm{P.~C.}~\sur{Canfield}}
	
	\author*[1,2]{\fnm{R.~J.}~\sur{McQueeney}}
	\email{mcqueeney@ameslab.gov}
	
	\author*[1,2]{\fnm{B.~G.}~\sur{Ueland}}
	\email{bgueland@ameslab.gov}
	
	\affil*[1]{\orgdiv{Division of Materials Sciences and Engineering}, \orgname{Ames National Laboratory, U.S. DOE, Iowa State University}, \orgaddress{ \city{Ames}, \postcode{50010}, \state{IA}, \country{USA}}}
	
	\affil[2]{\orgdiv{Department of Physics and Astronomy}, \orgname{Iowa State University}, \orgaddress{ \city{Ames}, \postcode{50010}, \state{IA}, \country{USA}}}
	
	\affil[3]{\orgdiv{Neutron Scattering Division}, \orgname{ Oak Ridge National Laboratory}, \orgaddress{ \city{Oak Ridge}, \postcode{37831}, \state{TN}, \country{USA}}}
	
	\affil[4]{\orgdiv{University of Missouri Research Reactor}, \orgname{ University of Missouri}, \orgaddress{ \city{Columbia}, \postcode{65211}, \state{MO}, \country{USA}}}
		
	\affil[5]{\orgdiv{Department of Chemistry}, \orgname{ University of Missouri}, \orgaddress{ \city{Columbia}, \postcode{65211}, \state{MO}, \country{USA}}}

	\date{\today}
	
	\abstract{Predicting magnetic ordering in kagome compounds offers the possibility of harnessing topological or flat-band physical properties through tuning of the magnetism.  Here, we examine the magnetic interactions and phases of ErMn$_6$Sn$_6$ which belongs to a family of $R$Mn$_6$Sn$_6$, $R=$~Sc, Y, Gd--Lu, compounds with magnetic kagome Mn layers, triangular $R$ layers, and signatures of topological properties. Using results from single-crystal neutron diffraction and mean-field analysis, we find that ErMn$_6$Sn$_6$ sits close to the critical boundary separating the spiral-magnetic and ferrimagnetic ordered states typical for nonmagnetic versus magnetic $R$ layers, respectively. Finding interlayer magnetic interactions and easy-plane Mn magnetic anisotropy consistent with other members of the family, we predict the existence of a number of temperature and field dependent collinear, noncollinear, and noncoplanar magnetic phases. We show that thermal fluctuations of the Er magnetic moment, which act to weaken the Mn-Er interlayer magnetic interaction and quench the Er magnetic anisotropy, dictate magnetic phase stability. Our results provide a starting point and outline a multitude of possibilities for studying the behavior of Dirac fermions in $R$Mn$_6$Sn$_6$ compounds with control of the Mn spin orientation and real-space spin chirality.}
		
	\maketitle
	
	\section{Introduction}

    The kagome lattice has a two-dimensional corner-sharing triangular arrangement which supports frustrated electronic interactions. Its electronic band structure generally contains both flat bands and linear (Dirac-like) band crossings, both of which can lead to important correlated electronic phenomena; flat bands can give rise to itinerant magnetic correlations and Stoner-type magnetism, whereas linear band crossings create Dirac cones and the associated topological (Dirac) fermions \cite{Yin_2022, Yang_2012, Tokura_2019, Ma_2021, Regnault_2022}. These latter features can lead to topological phenomena that intimately link charge and spin, such as dissipationless spin-momentum-locked (chiral) charge transport and quantum-anomalous-Hall effects \cite{Bernevig_2022}.
    
    Kagome compounds with competing interlayer magnetic interactions are particularly important as they can offer the ability to tune the electronic states of the kagome layers by manipulating the magnetism. Hexagonal $R$Mn$_6$Sn$_6$ ($R166$) metals with $R=$~Sc, Y, Gd--Lu, fall into this category. These compounds have emerged as model layered systems for magnetic tuning due to the combination of largely defect-free magnetic Mn kagome nets hosting topological electronic bands and $R$-site magnetism that controls details of the magnetic order \cite{Ma_2021, Yin_2020, Zhang_2020,Ghimire_2020,Dally_2021,Li_2021,Riberolles_2022, Lee_2023, Riberolles_2023, Annaberdiyev_2023, Lee_2024}. Crucial to harnessing the interplay between the magnetic order and topological properties is understanding the microscopic magnetic interactions determining the magnetic order and how to manipulate or influence the magnetic interactions to create a desired magnetic state. Here, we address these challenges via a single-crystal neutron diffraction and mean-field analysis study of ErMn$_6$Sn$_6$.

    As shown in Fig.~\ref{Fig:Struc}a for Er$166$, $R166$ compounds feature Mn kagome bilayers separated by triangular $R$ layers.  Importantly, the magnetic layers individually exhibit ferromagnetic (FM) ordering of their magnetic moments (spins), and a majority of $R166$ compounds with magnetic $R$ ions exhibit collinear ferrimagnetic (FIM) order driven by strong antiferromagnetic (AFM) coupling between the Mn and $R$ layers \cite{Baranov_2011}. For nonmagnetic $R$, such as Y, long range and competing Mn-Mn interlayer couplings lead to spiral magnetic order \cite{Venturini_1996, Rosenfeld_2008, Ghimire_2020}, and applied magnetic fields lead to distorted and noncoplanar spin configurations where the topological-Hall effect and planar-anisotropic magnetoresistance have been observed \cite{Ghimire_2020,Dally_2021,Wang_2022,Liu_2023}.  Thus, the strength of the Mn-$R$ coupling controls the stability between collinear and noncollinear magnetism in the $R$166 compounds and we show in this work that Er$166$ sits close to the critical boundary between the two types of ordering.

    Another key aspect of $R166$ compounds is that $R$ ions with nonzero orbital angular momentum adopt complex magnetic anisotropy through the crystalline-electric-field (CEF) splitting of their $4f$ orbitals.  Easy-plane Mn anisotropy is present, but strong $R$ anisotropy will dictate the Mn magnetization direction at low temperature and zero magnetic field. For the FM Mn layers, a uniaxial anisotropy is proposed to dramatically enhance the spin-orbit splitting of Dirac cones, forming a Chern insulator \cite{Yin_2020}. Further, the competition of $R$ anisotropy with the easy-plane Mn anisotropy can result in temperature and field driven spin-reorientation transitions between easy-plane and uniaxial ($R=$~Tb) or tilted ($R=$~Dy, Ho) FIM states. These transitions can facilitate magnetization switching via small changes in temperature or magnetic field \cite{Riberolles_2023, Malaman_1999,Jones_2022} which can create or destroy a Chern gap \cite{Yin_2020} or Weyl nodes \cite{Ghimire_2019}.  We find that the Er$^{3+}$ ion has weakly uniaxial and strongly temperature dependent anisotropy in Er$166$, resulting in facile magnetic switching. Small magnetic fields of $\mu_0H < 1$~T can trigger spin-reorientation transitions where large anomalous-Hall and topological-Hall responses have been reported \cite{Dhakal_2021}.
 
   Here, we present single-crystal neutron diffraction and mean-field analysis results that illustrate how competition between interlayer magnetic interactions and magnetic anisotropy in Er$166$ leads to a number of nearly degenerate magnetic states. Using this information, we characterize and understand the temperature driven first-order transition from planar-FIM to a distorted-triple-spiral order upon warming through $T_{\text{spiral}} = 92(1)$~K, showing that spins in the Er layers remain magnetically ordered above $T_{\text{spiral}}$ rather than their previously reported paramagnetic behavior \cite{Dhakal_2021}.  Using additional results from inelastic neutron scattering and magnetization measurements, we determine the key microscopic magnetic interactions and predict the emergence of various collinear, noncollinear, and noncoplanar magnetic phases as functions of temperature and magnetic field. Importantly, we find that the stability of the magnetic phases is controlled by thermal fluctuations of the Er spins which act to weaken the effective Mn-Er interlayer magnetic interaction and quench the Er magnetic anisotropy. 
	\section{Results and Discussion}	
	\subsection{Experimental Results}

	The magnetization $M$ of Er$166$ for a weak magnetic field of $\mu_0H=0.005$~T applied perpendicular to the crystalline $c$ axis is shown in Fig.~\ref{Fig:Chi_Diff}a and the zero-field neutron diffraction pattern for $(0,0,l)$ reciprocal-lattice points is shown in Fig.~\ref{Fig:Chi_Diff}b for increasing temperature. The main features in both figures agree with previously published data \cite{Malaman_1999,Suga_2006}; with cooling, a peak in $[M/H](T)$ in Fig.~\ref{Fig:Chi_Diff}a at the N\'eel temperature of $T_{\text{N}}=348(1)$~K coincides with the emergence of satellite magnetic-Bragg peaks in Fig.~\ref{Fig:Chi_Diff}b surrounding the $(0,0,2)$ structural-Bragg peak; a large jump in $[M/H](T)$ at $T_{\text{spiral}}$ accompanies the disappearance of the satellite peaks. The primary satellites [those satellite peaks closest to (0,0,2)] correspond to a temperature dependent AFM propagation vector of $\bm{\tau}=(0,0,\tau)$ which indicates the presence of magnetic ordering with a temperature dependent structure that is modulated along $\mathbf{c}$. Analysis of satellites for different values of the $h$ and $l$ reciprocal-lattice coordinates indicates that the spins primarily lie in the crystalline $\mathbf{ab}$ plane. Second-harmonic ($2\tau$) and third-harmonic ($3\tau$) satellites are also present in the figure and are addressed below.
	
	Figure~\ref{Fig:Chi_Diff}c shows that the integrated intensity of the $(1,0,\bar{\tau})$ primary satellite monotonically increases while cooling down to $T_{\text{spiral}}$ and that a jump in the integrated intensity of the $(1,0,0)$ Bragg peak accompanies the disappearance of the satellite at $T_{\text{spiral}}$.  This jump signals the emergence of a magnetic-Bragg peak on top of the $(1,0,0)$ structural-Bragg peak and is due to a transition to planar-FIM order with the spins continuing to lie in the $\mathbf{ab}$ plane (FIM-ab). The sharp jump and temperature hysteresis of the integrated intensity of the $(1,0,0)$ peak around $T_{\text{spiral}}$ are both consistent with a first-order transition. Figure~\ref{Fig:taus}d shows that the lattice parameters experience a small change in slope at $T_{\text{spiral}}$ but there is no conclusive sign of an accompanying structural transition.
	
\begin{figure}
	\centering
	\includegraphics[width=1\linewidth]{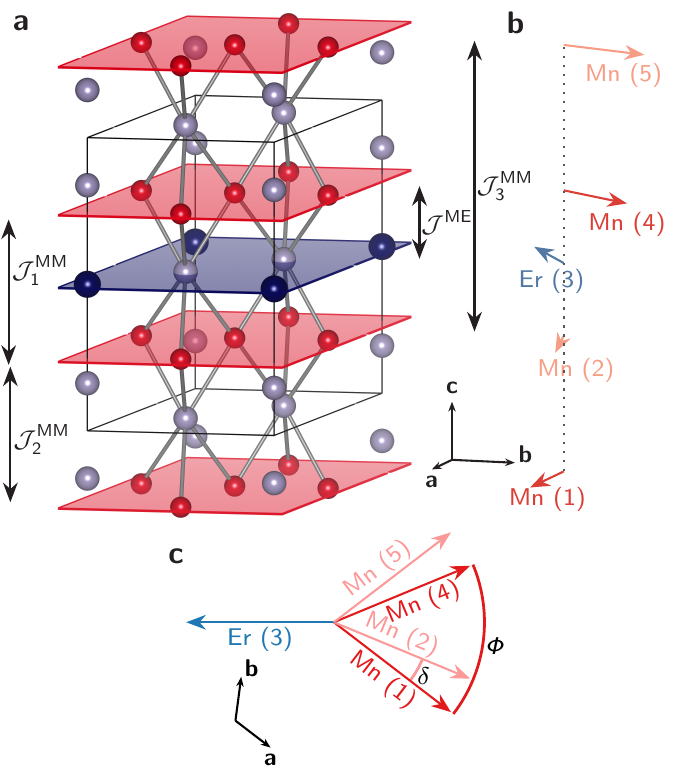}
	\caption{  \label{Fig:Struc} \textbf{Ideal triple-spiral magnetic ordering.} \textbf{a} Chemical structure of ErMn$_6$Sn$_6$ with the unit cell indicated by thin lines. 	The hexagonal $R166$ compounds crystallize in the HfFe$_6$Ge$_6$-type structure (space group $P6/mmm$, No.\ $191$) with lattice parameters of $a=5.51$~\AA\ and $c=9.00$~\AA\ at room temperature~\cite{Baranov_2011, Olenitch}. The individually ferromagnetic kagome-Mn  (triangular-Er) planes are indicated in red (blue).  Interlayer Mn-Sn bonds are displayed and interlayer-exchange interactions between Mn layers ($\mathcal{J}^{\text{MM}}_k$) and Mn and Er layers ($\mathcal{J}^{\text{ME}}$) are shown as black double arrows. \textbf{b} Parallel and \textbf{c} top-down views of ideal-triple-spiral order. Numbers label the sequence of the Mn and Er planes along $\mathbf{c}$ and each arrow represents a ferromagnetic Er or Mn plane.  Each plane has an ordered magnetic moment oriented perpendicular to $\mathbf{c}$ and the moment orientation varies between layers as shown. The angles $\Phi$ and $\delta$ characterize the magnetic order, with the Er moment direction pointing antiparallel to the bisector of $(\Phi-\delta)$. $\Phi=\delta=0$ corresponds to the low-temperature ferrimagnetic phase with the Er and Mn moments pointing opposite to each another.}
	\end{figure}

\begin{figure}
	\centering
	\includegraphics[width=1\linewidth]{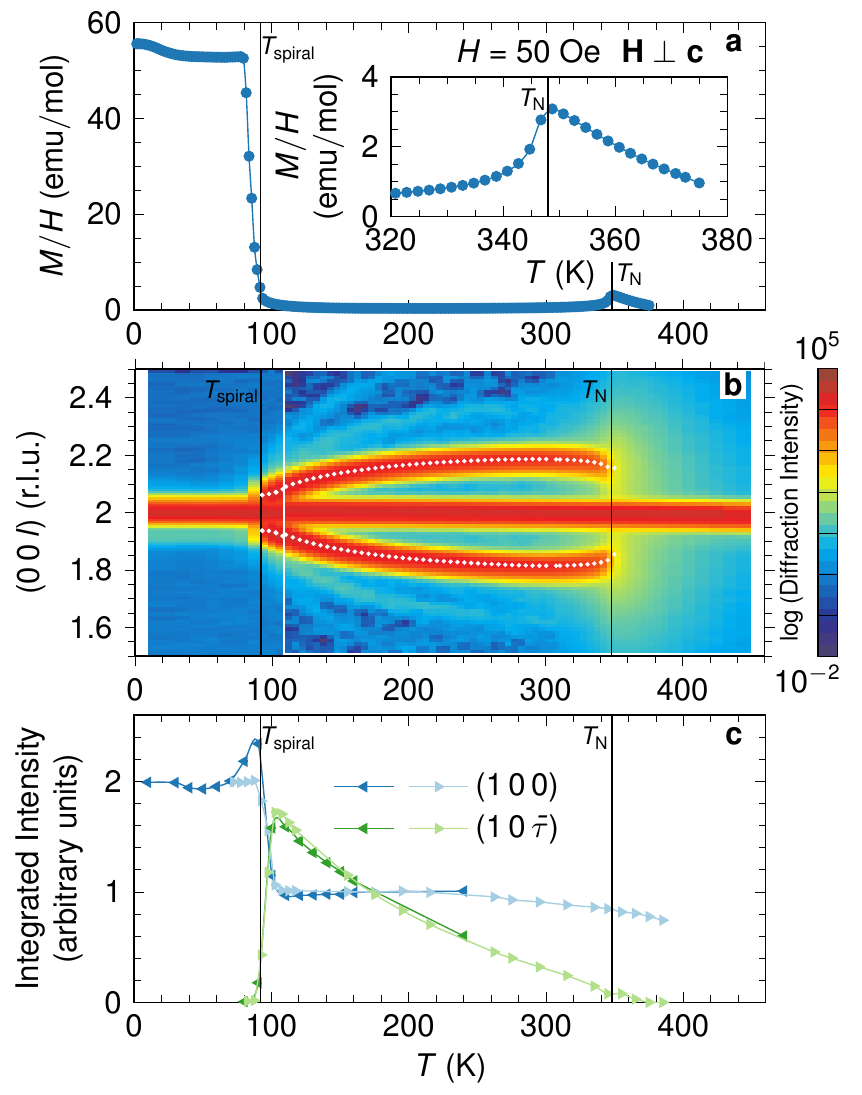}
	\caption{  \label{Fig:Chi_Diff} \textbf{Temperature-driven distorted-triple-spiral to ferrimagnetic ordering.} \textbf{a} Magnetization divided by field versus temperature for a magnetic field of $\mu_0H=0.005$~T applied perpendicular to the $\mathbf{c}$ crystalline axis. The inset shows a zoomed-in view of the high-temperature peak. \textbf{b} Diffraction pattern for  $(0,0,l)$ reciprocal-lattice points and increasing temperature.  Data for $T<105$~K and $T>105$~K are from two different experiments. A log scale is used for the intensity and white circles are the fitted centers of the $(0,0,2\pm\tau)$ magnetic-Bragg peaks. Similar data for $(1,0,l)$ are given in Supplementary~Section~1.1. \textbf{c} The temperature evolution of the integrated intensities of the $(1,0,0)$ and $(1,0,\bar{\tau})$ Bragg peaks. Darker (lighter) symbols correspond to measurements made on cooling (warming).}
	\end{figure}

\begin{figure*}
	\centering
	\includegraphics[width=1\linewidth]{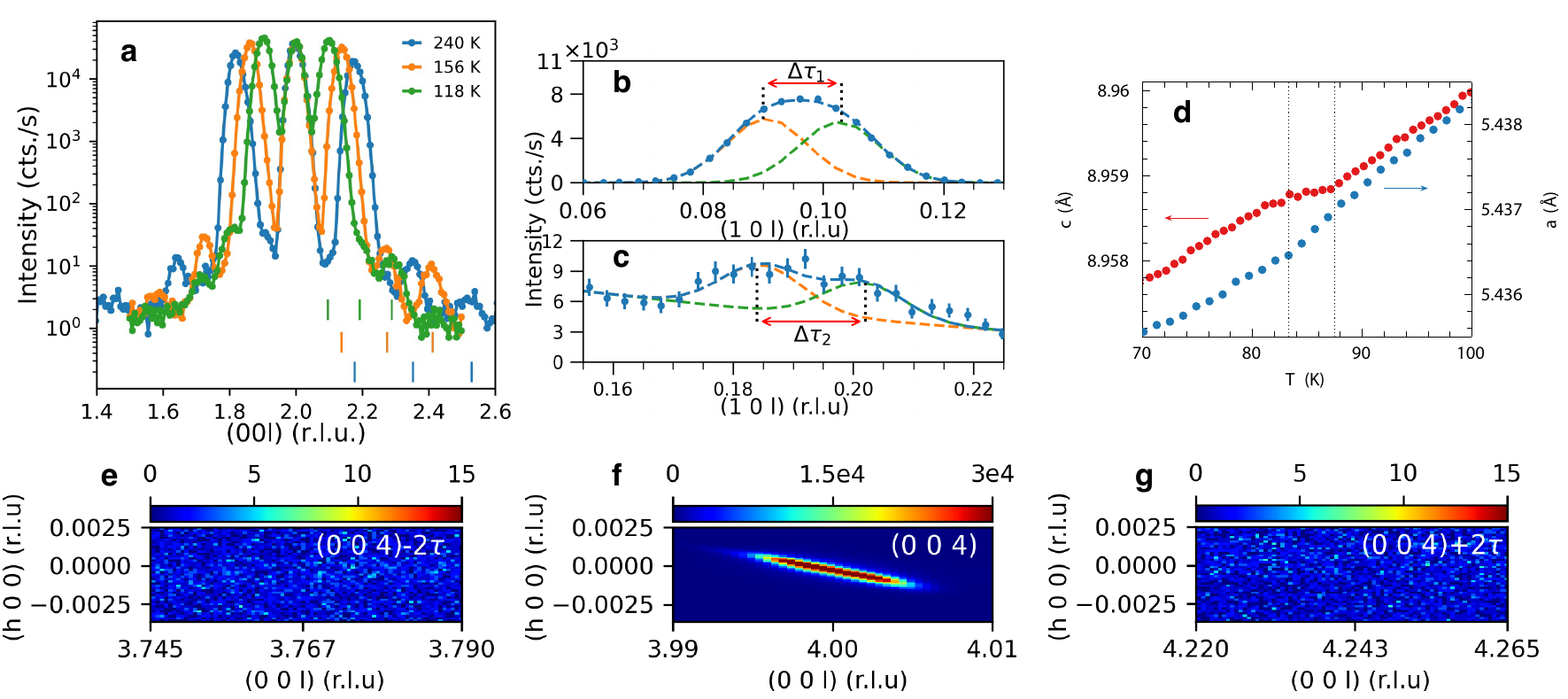}
	\caption{ \textbf{Higher harmonic magnetic-Bragg peaks.} \textbf{a} Diffraction patterns along $(0,0,l)$ for the indicated temperatures demonstrating the presence of higher-harmonic satellite peaks. Vertical lines indicate the centers  of the $\tau$, $2\tau$, and $3\tau$ peaks. \textbf{b},\textbf{c} Diffraction patterns showing the splitting of the lineshapes of the $(1,0,\tau)$ \textbf{b} and $(1,0,2\tau)$ \textbf{c} satellite peaks. Dashed green and orange lines show individual gaussian components of the lineshapes and vertical dotted lines indicate the center of each component. $\Delta\tau_1$ and $\Delta\tau_2$ show the distances between the centers of the components. \textbf{d} The $a$ and $c$ lattice parameters plotted versus temperature as determined from single-crystal x-ray diffraction. \textbf{e}--\textbf{g} Single-crystal x-ray diffraction data for $T=140$~K taken across the $(0,0,4-2\tau)$ \textbf{e},  $(0,0,4)$ \textbf{f}, and $(0,0,4+2\tau)$ \textbf{g} positions. The color bar scales are counts per second. The absence of Bragg peaks in \textbf{e} and \textbf{g} is evidence that the $2\tau$ peaks are not structural in origin. Error bars indicate one standard deviation.
		\label{Fig:taus}}
	\end{figure*}

     Previous reports attributed the transition at $T_{\text{spiral}}$ to a loss of Er magnetic ordering for $T> T_{\text{spiral}}$ \cite{Malaman_1999}, leaving a Mn only double-spiral structure similar to that found for Y$166$ \cite{Rosenfeld_2008, Ghimire_2020, Dally_2021}. The double-spiral order is described by a small angle $\delta$ between the spin directions of the Mn bilayers [e.g.\ layers ($1$) and ($2$) in Figs.~\ref{Fig:Struc}b and \ref{Fig:Struc}c], which are strongly coupled via the exchange interaction $\mathcal{J}^{\text{MM}}_{2}$ indicated in Fig.~\ref{Fig:Struc}a. A larger angle $\Phi=2\pi\tau$ describes a compound rotation of the spins between bilayers and encapsulates the periodicity of the spiral. For Er$166$, however, the temperature dependence of $\tau$ that we observe is much stronger than it is for Y$166$ \cite{Ghimire_2020,Wang_2022} and is similar to observations for the triple-spiral order of TmMn$_6$Sn$_{5.8}$Ge$_{0.2}$ \cite{Lefevre_2002}. In addition, we find that if we assume double-spiral order, the integrated intensity of the $(1,0,\bar{\tau})$ satellite would give a Mn ordered magnetic moment of $\mu_{\text{Mn}}\approx8~\mu_B$ at $T=108$~K. This value is too large for Mn-only magnetic ordering. Furthermore, as described below, the temperature dependence of the integrated intensity is in reasonable agreement with the mean-field analysis which predicts triple-spiral magnetic order.

	The ideal triple-spiral magnetic order is similar to double-spiral order but with the Er spins participating in the ordering and pointing antiparallel to the bisector of $(\Phi-\delta)$. This is shown in Figs.~\ref{Fig:Struc}b and  \ref{Fig:Struc}c.  The spins of the FM Er layer rotate in phase with the Mn spiral order, with the total triple spiral having a period of $c/\tau$. As we show below, the mean-field results in combination with the single-crystal neutron diffraction data strongly support the presence of triple-spiral ordering between $T_{\text{N}}$ and $T_{\text{spiral}}$. We also have performed a Rietveld refinement using \textsc{fullprof} \cite{fullprof} to powder neutron diffraction data taken at $T=200$~K.  Supplementary~Section~1.2 shows that the refinement yields good agreement with the triple-spiral structure, with $\tau=0.1876(6)$ and $\delta=14.0(2)$\degree, and reasonable values for the ordered Er and Mn magnetic moments of $\mu_{\text{Er}}=3.9(3)~\mu_B$ and  $\mu_{\text{Mn}}=2.0(1)~\mu_B$, respectively. 
  
\begin{figure*}
	\centering
	\includegraphics[width=1\linewidth]{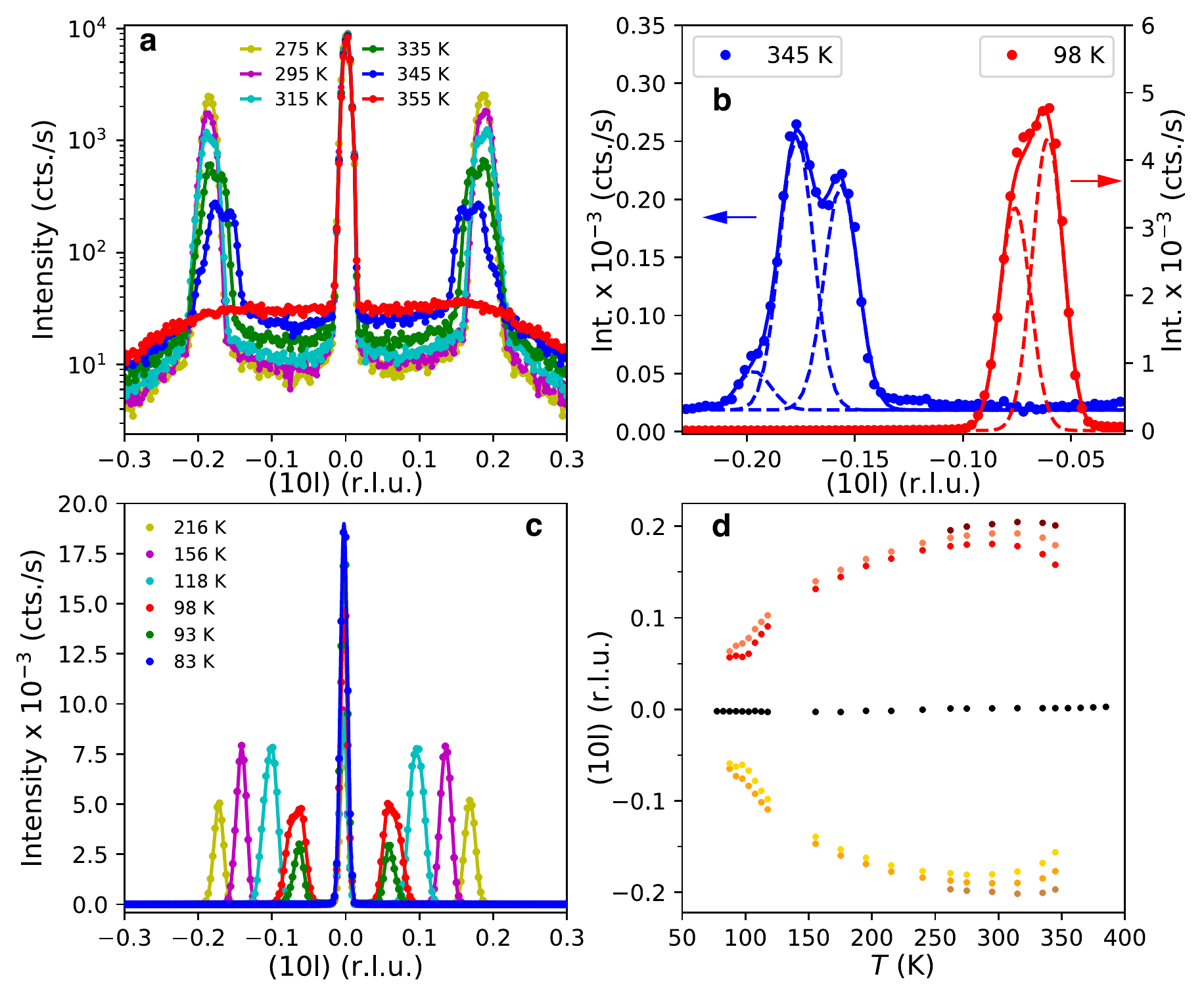}
	\caption{ \textbf{Temperature-dependent splitting of the magnetic-Bragg peaks.} \textbf{a} $(1,0,l)$ neutron diffraction patterns for temperatures crossing the N\'eel temperature of $T_{\text{N}}=348(1)$~K. \textbf{b} The $(1,0,\bar{\tau})$ magnetic-Bragg peak for $T=345$ and $98$~K. Gaussian components of the lineshapes are shown by dashed lines, where the higher temperature lineshape has three gaussian components and the lower temperature lineshape contains two gaussian components. \textbf{c} $(1,0,l)$ neutron diffraction patterns for temperatures within the triple-spiral ($T>T_{\text{spiral}}$) and ferrimagnetic-ab ($T<T_{\text{spiral}}$) phases, where $T_{\text{spiral}}=92(1)$~K. \textbf{d} Temperature evolution of the center of the $(1,0,1)$ Bragg peak and the centers of the gaussian components of the lineshapes for the $(1,0,\pm\tau)$ magnetic-Bragg peaks.  Error bars indicate one standard deviation.   
		\label{Fig:splitting}}
	\end{figure*}

	In addition to the temperature dependent primary satellites, two other remarkable features appear in the neutron diffraction data. First, much weaker satellite peaks are evident in Fig.~\ref{Fig:Chi_Diff}b that are $2\bm{\tau}$ and $3\bm{\tau}$ away from $(0,0,2)$. These higher-harmonic satellite peaks are highlighted in the constant-temperature cuts in Fig.~\ref{Fig:taus}a.  The second remarkable feature is the splitting of the lineshapes of individual satellites, as shown in Fig.~\ref{Fig:splitting}. Both of these features suggest that the triple-spiral order is distorted in Er$166$ and we next address each feature in turn.
 
    Odd-harmonic magnetic satellites are usually associated with squaring up of the AFM order \cite{Jayasekara_2014} (such as ``bunched-spiral'' order) whereas it is much less common to observe even-harmonic magnetic satellites.  Possible sources for even-harmonic satellites include distortions of the chemical lattice, magnetoelastic distortions of the ideal triple-spiral magnetic order, or the existence of fan-type magnetic order similar to that which occurs when applying an in-plane magnetic field to a magnetic spiral \cite{Johnston_2017, Johnston_2017E, Johnston_2019}. Figures~\ref{Fig:taus}e to \ref{Fig:taus}g show that we do not observe the $2\tau$ peaks in single-crystal x-ray diffraction data, which suggests that the $2\tau$ satellites arise from magnetic diffraction.
    
    Regarding the possibility of fan-type magnetic order, we hypothesize that a net magnetization must exist within a magnetic domain in order for it to occur with no applied magnetic field. This could happen if the orientation of the Er spins is pinned by a strong CEF potential. However, both our neutron diffraction and magnetization measurements do not find evidence for a net magnetization existing between $T_{\text{spiral}}$ and $T_{\text{N}}$, as we do not conclusively measure any magnetic contributions at integer $(h,k,l)$ positions in the diffraction data and zero-field-cooled and field-cooled magnetization measurements made with $\mu_0H=0.1$~T do not reveal a domain population imbalance. Our mean-field modeling, described below, is also dismissive of zero-field fan-type order. Thus, we associate the $2\tau$ peaks with an unknown distortion of the ideal triple-spiral magnetic order.
 
	Next, the splitting of the lineshapes of the magnetic-Bragg peaks is also suggestive of a distortion of the ideal triple spiral.  The cuts in Figs.~\ref{Fig:splitting}a to \ref{Fig:splitting}c show that the lineshape of a primary-satellite peak is not described by a gaussian lineshape.  Rather, depending on temperature, the lineshape is better fit by the sum of two ($T\le 256$~K) or three ($T>256$~K) gaussian peaks and a background.  This is exemplified in Fig.~\ref{Fig:splitting}b for the $(1,0,\bar{\tau})$ primary satellite at $345$~K and $98$~K where the individual gaussian components are indicated by dashed lines.
	
	A similar splitting of the primary-satellite lineshape has been reported for Y$166$ \cite{Venturini_1996,Ghimire_2020,Dally_2021} and Ga-substituted Y$166$ \cite{Xu_2021, Bhandari_2023}, with the splitting for Y$166$ being characterized by $\Delta\tau_1 = \tau^+-\tau^- \approx 0.05$ at room temperature.  Here, $\tau^+$ and $\tau^-$ refer to the centers of the two gaussian components contributing to the lineshape. As shown in Figs.~\ref{Fig:splitting}b and \ref{Fig:splitting}d, we find that three rather than two gaussian components contribute to the lineshapes of the $(1,0,\pm\tau)$ primary satellites of Er$166$ around room temperature, with a somewhat smaller splitting of $\Delta\tau_1 \approx 0.03$ between components. Similar to Y$166$, Fig.~\ref{Fig:splitting}d shows that the splitting is reduced with decreasing temperature, becoming barely resolvable at intermediate temperatures. However, further cooling results in an increase of the splitting where the third gaussian component never reemerges. 

    Like the double-spiral ordering of Y$166$ \cite{Ghimire_2020}, the origin of the splitting of the primary-satellite lineshapes for the triple-spiral ordering of Er$166$ is not yet completely clear. We show in Supplementary~Sections~1.2 and 1.3 that the lineshape splitting is observed for different single-crystal samples and even in the powder neutron diffraction data. This points to the splitting being an intrinsic property. The lineshape splitting could arise from different magnetic domains with slightly different spiral periods or it could result from beating of the spiral order within a single magnetic domain that results in long-period ($\propto 1/\Delta\tau_1$) modulations of the magnetic structure. The single-domain origin of the lineshape splitting gains some credence from the observed broadening of the $2\tau$ and $3\tau$ satellites, which is consistent with splitting of their lineshapes. However, Figs.~\ref{Fig:taus}b and \ref{Fig:taus}c surprisingly show that the splittings of the $(1,0,\tau)$ and $(1,0,2\tau)$ lineshapes are unequal, with $\Delta\tau_2 \approx 1.5\Delta\tau_1$.  Though not completely understood, this observation provides strong evidence supporting the magnetic nature of the $2\tau$ satellites and lends weight to the single-domain hypothesis for the lineshape splitting. 

\subsection{Magnetic Hamiltonian}
	
	The competition between exchange coupling and magnetic anisotropy makes the development of a microscopic magnetic model for Er$166$ challenging, but necessary to understand the magnetic phase stability. We define a magnetic Hamiltonian comprised of isotropic-exchange interactions ($\mathcal{H}_{\text ex}$) between Mn-Er and Mn-Mn spins, a Zeeman term ($\mathcal{H}_{\text Z}$) for coupling spins to an externally applied magnetic field, and single-ion terms for the Er ($\mathcal{H}_{\text{Er}}$) and Mn ($6\mathcal{H}_{\text{Mn}}$) crystallographic sites that capture the magnetic anisotropy.
 
    The exchange Hamiltonian is given by
	\begin{equation}
		\mathcal{H}_{\text ex}=\sum_k\sum_{i<j} \mathcal{J}^{\text{MM}}_{k} \mathbf{s}_i \cdot \mathbf{s}_j + \mathcal{J}^{\text{ME}} \sum_{\langle i<j \rangle} \mathbf{s}_i \cdot \mathbf{S}_j\,,
		\label{Hex}
	\end{equation}
	where we label various intralayer and interlayer interactions between Mn spins ${\mathbf s}_i$ as $\mathcal{J}^{\text{MM}}_{k}$ where $k$ is a layer index. $\mathcal{J}^{\text{ME}}>0$ is the AFM coupling between neighboring Mn and Er spins ${\mathbf S}_j$, and all of the pertinent interlayer interactions are diagrammed in Fig.~\ref{Fig:Struc}a.  The values of $\mathcal{J}^{\text{MM}}_k$ are similar to those found for Tb$166$ \cite{Riberolles_2022}, and whereas the intralayer Mn-Mn interaction $\mathcal{J}^{\text{MM}}_{0}$ is large and FM, it is known that the frustrated Mn-Mn interlayer ($k>0$) interactions lead to competition between collinear and spiral phases \cite{Baranov_2011, Rosenfeld_2008}. The Zeeman energy is given by
	\begin{equation}
		\mathcal{H}_{\text{Z}} = -\mu_{\text B}(6g{\mathbf s} + g_J{\mathbf J}) \cdot \mu_0{\mathbf H}
		\label{eqn:Zeeman}
	\end{equation}
	where $g \approx 2$ for Mn, and $g_J=6/5$, $J=15/2$  for Er, giving $g_JJ=9$.

Complex behavior of the magnetic anisotropy arises due to the action of the CEF potential of neighboring ions on the $4f$ orbital states of Er. The Hamiltonian for the CEF acting on Er is
	\begin{equation}
		\mathcal{H}_{\text{Er}}=B_2^0 \mathcal{O}_2^0+B_4^0 \mathcal{O}_4^0+B_6^0 \mathcal{O}_6^0+B_6^6 \mathcal{O}_6^6
		\label{eqn:Er_Ham}
	\end{equation}
where $B_l^m$ are the CEF parameters for Er$^{3+}$ with hexagonal point-group symmetry and $\mathcal{O}_l^m$ are Stevens operators.  We find that our CEF parameters  lead to uniaxial magnetic anisotropy for Er in the ground state, although barely so. The uniaxial Er anisotropy competes with the simple Mn easy-plane anisotropy given by
	\begin{equation}
		\mathcal{H}_{\text{Mn}}=K^{\text{M}}s_z^2
		\label{eqn:Mn_Ham}
	\end{equation}
	where $K^{\text{M}}>0$. Easy-plane Mn anisotropy is consistent with the planar-FIM or helical ground states found in $R166$ compounds with magnetically isotropic ($R=$~Gd) or non-magnetic ($R=$~Sc, Y, Lu) ions, respectively \cite{Venturini_1996,Malaman_1999}.
 
    A representative set of parameters for the magnetic Hamiltonian are shown in Table~\ref{tbl:params} and their estimation is described in Supplementary~Section~1.4.   Using these parameters, we next describe a mean-field analysis of the equilibrium states which provides semi-quantitative agreement with observations of the temperature driven FIM to triple-spiral transition and field and temperature driven spin-reorientation and spin-flop transitions. The strong intralayer Mn-Mn exchange justifies the assumption that each Mn and Er layer remains FM upon cooling or upon the application of a magnetic field, but both the direction and magnitude of the spins can vary from layer to layer. Thus, determination of the magnetic structure can require minimization of the free energy for spin configurations potentially spanning dozens of unit cells.  To facilitate the calculations, we initially ignore planar anisotropy of the Er ion [i.e. the $B^6_6$ term in Eqn.~\eqref{eqn:Er_Ham}] and consider only uniaxial applied fields.
    
    Within a mean-field description, the magnetic structure is determined by minimizing the magnetic free energy described by four angles $\theta_{\text{Er}}$, $\theta_{\text{Mn}}$, $\Phi$, and $\delta$. $\theta_{\text{Er}}$ ($\theta_{\text{Mn}}$) describes the polar angle of the Er (Mn) spins away from $\mathbf{c}$, and $\Phi$ and $\delta$ are the spiral angles defined above. Minimization of the $\mathcal{J}^{\text{ME}}$ exchange energy requires that $\bm{\mu_{\text{Er}}}$ points antiparallel to the bisector of $(\Phi-\delta)$, as shown in Fig.~\ref{Fig:Struc}c. With these constraints, the possible magnetic phases are planar-FIM (FIM-ab), uniaxial-FIM (FIM-c), vertical-plane-canted  (VP-canted), planar-spiral, vertical-conical-spiral (VCS), and forced-FM (FF). Further details of the mean-field calculations are given Supplementary~Sections~1.5 and 1.6.

	\subsection{Mean-Field Results and Discussion}
	
\begin{figure}
	\includegraphics[width=0.8\linewidth]{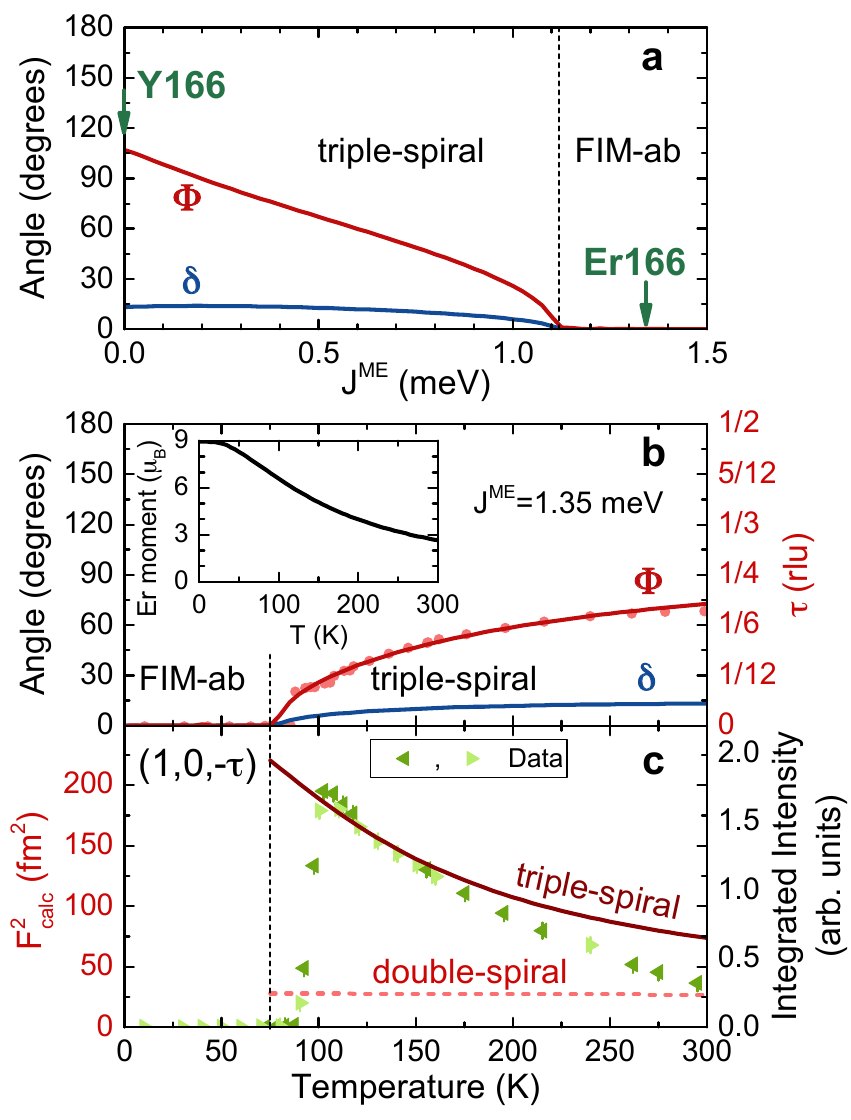}
	\caption{ \textbf{Mean-field analysis of the zero-field magnetic phases.} \textbf{a} The evolution from planar-ferrimagnetic (FIM-ab) to ideal-triple-spiral magnetic order as a function of the Mn-Er interlayer interaction $\mathcal{J}^{\text{ME}}$. The angles $\Phi$ and $\delta$ are defined in Fig.~\ref{Fig:Struc}\textbf{c}, and green arrows indicate the $T=0$~K magnetic phases for Y$166$ ($\mathcal{J}^{\text{ME}}= 0$) and Er$166$ ($\mathcal{J}^{\text{ME}}= 1.35$~meV). \textbf{b} The evolution from FIM-ab to ideal-triple-spiral order with temperature. Dots are the experimental values for the spiral periodicity from Fig.~\ref{Fig:Chi_Diff}b. The inset shows the reduction of the Er ordered magnetic moment with temperature due to thermal fluctuations. \textbf{c} The squares of the magnetic structure factors for ideal-triple-spiral and double-spiral order for the $(1,0,\bar{\tau})$ satellite peak calculated using the mean-field values for $\Phi$, $\delta$, and the Mn and Er ordered magnetic moments. The measured integrated intensity versus temperature for the $(1,0,\bar{\tau})$ satellite (left and right triangles) is also shown. Error bars indicate one standard deviation.}
	\label{fig:mean_field_xy}
	\end{figure}

	
	Competition between the weaker FM $\mathcal{J}^{\text{MM}}_1$ and AFM $\mathcal{J}^{\text{MM}}_3$ interactions lead to the zero-field spiral phases in the hexagonal $R166$s with non-magnetic $R=$~Sc, Y, or Lu \cite{Rosenfeld_2008}. For $R166$s with magnetic $R$ ions,  the $\mathcal{J}^{\text{ME}}$ coupling destabilizes the spiral state in favor of collinear-FIM order, as shown in Fig.~\ref{fig:mean_field_xy}a. The spiral phase is the preferred $T=0$~K state for $\mathcal{J}^{\text{ME}}\approx0$ and as $\mathcal{J}^{\text{ME}}$ is increased, $\Phi\rightarrow0$ and FIM-ab becomes the preferred ground state above a critical value of $\mathcal{J}^{\text{ME}} \approx 1.1$~meV.  Thus, the experimentally determined value of $\mathcal{J}^{\text{ME}}=1.35$~meV for Er$166$ correctly predicts the FIM-ab ground state. 
 
    We have performed mean-field calculations for $T>0$~K to test the stability of the FIM-ab phase of Er$166$ at higher temperatures. We find that with rising temperature the accompanying increase in thermal fluctuations of the Er spins destabilizes the FIM-ab phase in favor of the ideal triple spiral at $T_{\text{spiral}}$. This is exemplified by the plots in Fig.~\ref{fig:mean_field_xy}b which show that the ideal triple spiral becomes the equilibrium state above $T=75$~K, which is in good agreement with the experimentally observed value of $T_{\text{spiral}} = 92(1)$~K. Figure~\ref{fig:mean_field_xy}b also shows that the calculated temperature dependence of $\Phi$ aligns almost perfectly with the experimentally determined values of $\Phi$ from Fig.~\ref{Fig:Chi_Diff}b. These results give very strong confidence in the chosen interlayer-exchange parameters shown in Table \ref{tbl:params}. Finally, Supplementary~Figure~8 shows that only a small free energy difference exists between the FIM-ab and ideal triple-spiral phase. This is consistent with observations that the FIM-ab phase can be stabilized at $T>T_{\text{spiral}}$ by a very small planar magnetic field \cite{Canepa_2005}.
    
    Previous reports have ascribed the transition at $T_{\text{spiral}}$ to a decoupling of Mn and Er magnetic sublattices and the complete loss of Er ordering~\cite{Malaman_1999}.  Our mean-field and experimental results indicate that Er participates in the spiral, leading to the experimentally observed distorted-triple-spiral phase, although there is a substantial reduction of $\mu_{\text{Er}}$ with increasing temperature, as shown in the inset to Fig.~\ref{fig:mean_field_xy}b. Figure~\ref{fig:mean_field_xy}c shows the squares of the magnetic structure factors for the double-spiral and ideal triple-spiral orders calculated for the $(1,0,\bar{\tau})$ satellite peak using the mean-field determined parameters. The curve for the ideal triple spiral agrees well with the neutron diffraction data, showing concave-up behavior. The double-spiral curve, on the other hand, shows virtually no temperature dependence below $T = 300$~K.  From these comparisons, we also understand that the continued shortening of the triple-spiral period with increasing temperature is caused by growing thermal fluctuations of the Er spins which progressively reduce $\mu_{\text{Er}}$ and the effective Mn-Er coupling.

\begin{figure}[]
	\includegraphics[width=1.0\linewidth]{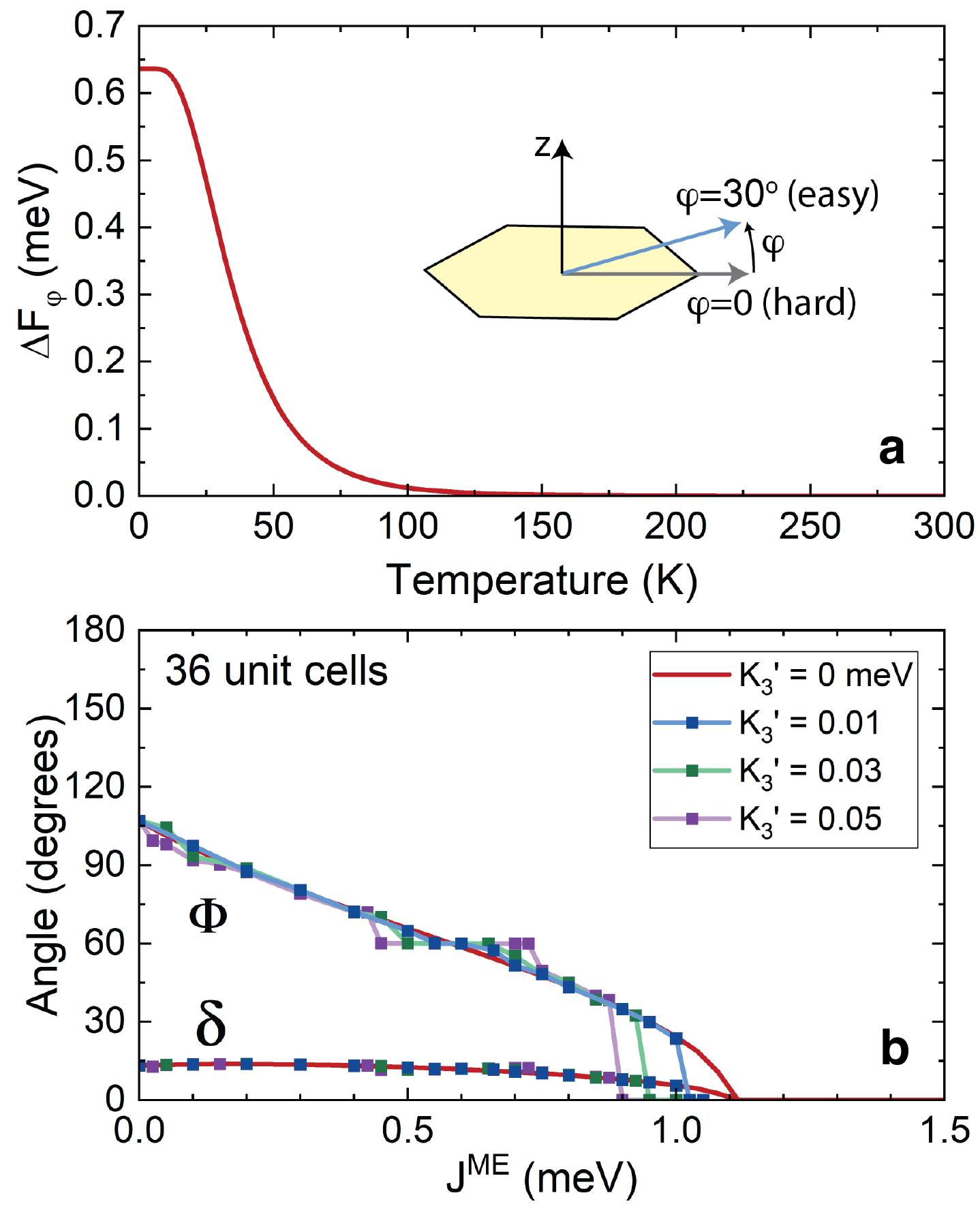}
	\caption{\textbf{Role of planar magnetic anisotropy in establishing the magnetic phases.} \textbf{a} Temperature evolution of the planar-magnetic-anisotropy energy estimated as the free-energy difference between easy-axis ($\varphi_{\text{Er}}=30$\degree, $\varphi_{\text{Mn}}=210$\degree) and hard-axis ($\varphi_{\text{Er}}=0$\degree, $\varphi_{\text{Mn}}=180$\degree) spin orientations for a FIM-ab phase in the mean-field approximation. \textbf{b} Average spiral angles adopted by the $T=0$~K minimum energy solution for planar spins in a stack of $36$ Mn-Er-Mn layers and for different values of the planar-anisotropy parameter $K_3^{\prime}$.  }
	\label{fig:planar_MAE}
\end{figure}

	For more insight into the microscopic details of the magnetism giving rise to a high temperature triple-spiral phase, we next consider the six-fold planar anisotropy term $B_6^6$ for the Er and whether its proper treatment leads to fan-like phases or distortions of the ideal triple-spiral order that generate the $2\tau$ and $3\tau$ satellite neutron diffraction peaks.  We begin by estimating the temperature dependence of the planar MAE by calculating the free-energy difference $\Delta\mathcal{F}_\varphi$ between FIM-ab phases with spins pointing either along the easy ($\varphi_{\text{Er}}=30$\degree, $\varphi_{\text{Mn}}=210$\degree) or hard ($\varphi_{\text{Er}}=0$\degree, $\varphi_{\text{Mn}}=180$\degree) planar axis. Figure~\ref{fig:planar_MAE}a shows that thermal fluctuations of the Er spins leads to a rapid decrease of $\Delta\mathcal{F}_\varphi$ with increasing temperature, with the planar MAE being reduced by $95$\% at $T_{\text{spiral}}$ ($\Delta\mathcal{F}_\varphi\approx0.035$~meV) and becoming negligible above $T=150$~K. 
	
	We next examine whether a certain size of the planar-MAE constant $K_3^{\prime}$, where $K_3^{\prime}=J^{(6)}B_6^6=\Delta\mathcal{F}_\varphi/2$, can lead to distorted-spiral order using results from classical energy-minimization calculations for $T=0$~K and a stack of $36$ Mn-Er-Mn layers. Figure~\ref{fig:planar_MAE}b shows that larger values of $K_3^{\prime}$ do not lead to fan-like phases, but for moderate values of $\mathcal{J}^{\text{ME}}$ larger values of $K_3^{\prime}$ lead to a lock-in of the spiral periodicity at $\Phi=60$\degree\ ($\tau=1/6$), where the six-fold planar MAE of the Er spins is fully minimized.  We also find that nonzero values of $K_3^{\prime}$ lead to a first-order like jump in the spiral periodicity at $T_{\text{spiral}}$, in agreement with observations. 
	
	It is difficult to address whether distortions of the ideal triple-spiral order close to $T_{\text{spiral}}$ can result from moderate values of $K_3^{\prime} < 0.02$ meV. Such distortions, for example, could lead to even and odd-harmonic satellite peaks caused by the pinning of the Er spin direction by the planar MAE.  However, three factors suggest that the planar anisotropy plays no role in producing the higher-harmonic satellites: ($1$) the higher-harmonic satellites persist well above $T=200$~K, where the Er-planar MAE is fully quenched; ($2$) there is no experimental evidence for a lock-in of the spiral periodicity to $60$\degree; ($3$) anharmonicity of the triple spiral resulting from Er spin bunching is expected to produce only odd-harmonic satellites \cite{Ramazanoglu_2011,Ratcliff_2016,Chaix_2016}.
 	
\begin{figure*}
	\includegraphics[width=1. \linewidth]{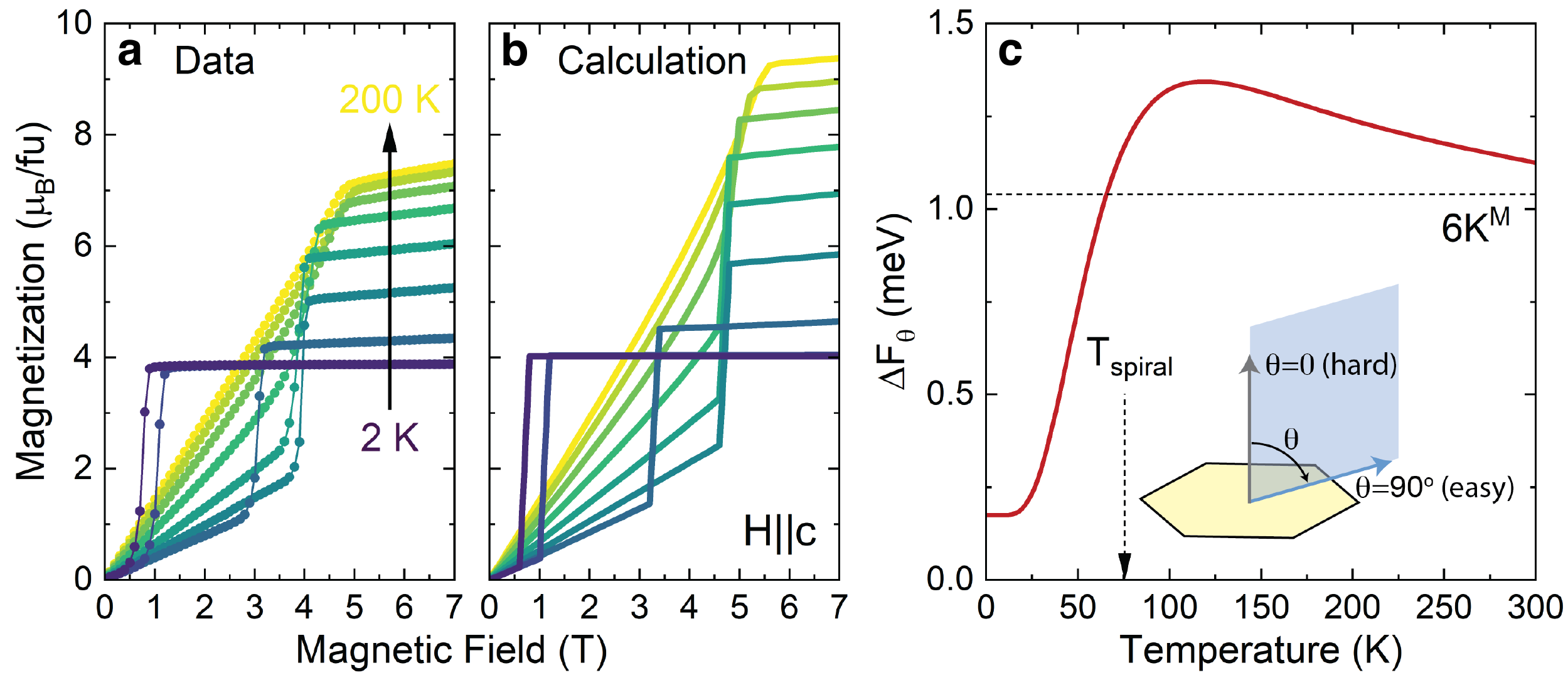}
	\caption{\textbf{Field-induced magnetic phases and the role of polar magnetic anisotropy.} \textbf{a} Measured magnetization versus magnetic field data for several temperatures with the field applied parallel to $\mathbf{c}$.  \textbf{b} Complementary results from mean-field calculations using the parameters in Table~\ref{tbl:params}.  \textbf{c} Results from mean-field calculations for the temperature evolution of the polar magnetic-anisotropy energy, which is estimated as the free-energy difference between uniaxial ($\theta_{\text{Mn}}=0$\degree, $\theta_{\text{Er}}=180$\degree) and planar ($\theta_{\text{Mn}}=\theta_{\text{Er}}=90$\degree) spin orientations for a ferrimagnetic phase. $K^{\text{M}}$ is the Mn anisotropy parameter. }
	\label{fig:polar_MAE}
	\end{figure*}

$M(H)$ data for $\mathbf{H}\parallel\mathbf{c}$ measured for various temperatures are shown in Fig.~\ref{fig:polar_MAE}a. Similar to previous results \cite{Suga_2006}, sharp steps to a plateau are visible for $T<T_{\text{spiral}}$.  These steps correspond to a first-order magnetization process (FOMP) that coherently rotates the planar spins of the FIM-ab phase to lay along $\mathbf{c}$, resulting in the FIM-c state.  The critical FOMP field is small at low temperature ($\mu_0H_{\text{FOMP}}=0.65$~T) and increases with increasing temperature to a maximum of $\approx4$~T close to $T_{\text{spiral}}$.
 
We understand the magnetization data using our mean-field model by calculating the polar MAE which we define as the free-energy difference $\Delta\mathcal{F}_\theta$ between FIM-ab ($\theta=90$\degree) and FIM-c phases ($\theta=0$\degree). Figure~\ref{fig:polar_MAE}c shows that the polar MAE is small at low temperatures, indicating the near degeneracy of FIM-ab and FIM-c phases caused by competing Er uniaxial and Mn easy-plane anisotropies. This is consistent  with the small value for $H_{\text{FOMP}}$ at $T=2$~K seen by experiment, and our mean-field calculations for $M(H)$ shown in Fig.~\ref{fig:polar_MAE}b are similar to the experimental data in Fig.~\ref{fig:polar_MAE}a, correctly predicting a small $H_{\text{FOMP}}$ for low temperatures.  Upon increasing the temperature, $H_{\text{FOMP}}$ shifts to higher fields, indicating that thermal fluctuations and quenching of the Er anisotropy increase the net easy-plane anisotropy. Above $T_{\text{spiral}}$, the steps in $M(H)$ broaden with increasing temperature until the curves show linear behavior, rather than a step, before plateauing.  This departure from the step-like jump signals an $(H,T)$ region where a gradual canting of the spins occurs with increasing field. The canting creates a VCS magnetic order which terminates in the FIM-c phase at higher field.  

The increase of $H_{\text{FOMP}}$ and disappearance of the FOMP above $T_{\text{spiral}}$ is again due to increasing thermal fluctuations of the Er spins with increasing temperature, which in this case results in a temperature-driven decrease of the contribution of the Er uniaxial anisotropy to the MAE. Surprisingly, the single-ion anisotropy of Er attains a planar character above $T\approx100$~K, resulting in the maximum in $\Delta\mathcal{F}_\theta$ shown in Fig.~\ref{fig:polar_MAE}c. At higher temperatures, Er contributions to the MAE become completely quenched by thermal fluctuations and the MAE approaches the classical value for the Mn ions of $6K^{\text{M}}=1.02$~meV.  At all temperatures, the mean-field calculations of $M(H)$ show qualitative agreement with the experimental curves which confirms the assignment of the FIM-ab, VCS, and FIM-c magnetic phases. This is despite a likely overestimation of the size of the Er spin thermal fluctuations by the calculations.

\begin{figure}
		\includegraphics[width=1. \linewidth]{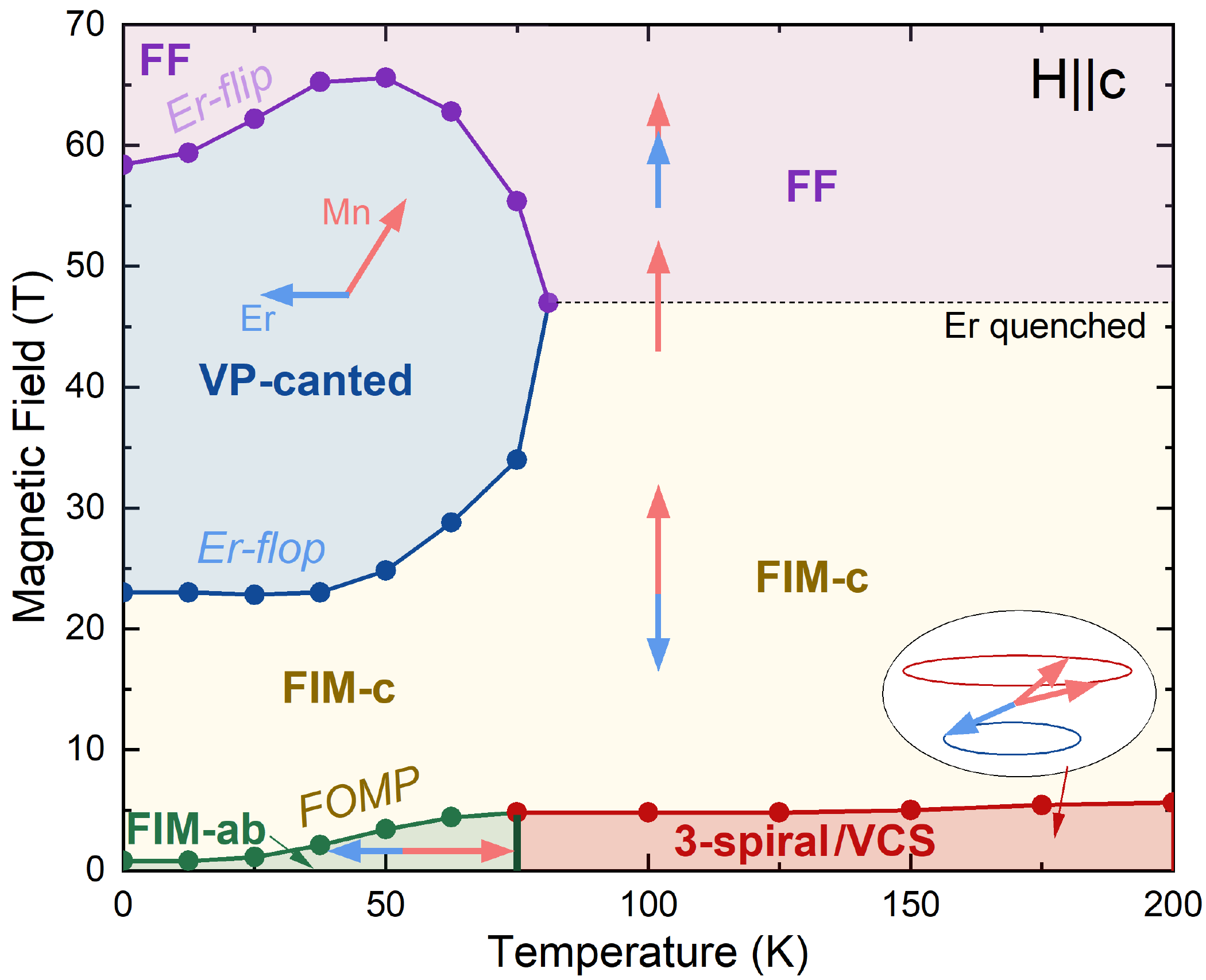}
		\caption{\textbf{$\mathbf{H} \parallel \mathbf{c}$ mean-field magnetic phase diagram.} The phases are: planar-ferrimagnetic (FIM-ab), uniaxial-ferrimagnetic (FIM-c), triple-spiral ($3$-spiral), vertical-conical-spiral (VCS), vertical-plane-canted (VP-canted), and forced-ferromagnetic (FF).  Metamagnetic transitions are labeled as: first-order magnetization process (\textit{FOMP}), Er-spin flop (\textit{Er-flop}), and Er-spin flip (\textit{Er-flip}).  The horizontal dotted line indicates the FIM-c to FF crossover where the ordered Er magnetic moment is completely quenched.  Red (blue) arrows show the orientation of the Mn (Er) spins in different layers.}
		\label{fig:phase_diagram}
	\end{figure}

	Given its success in describing the low-field phases, we employ the mean-field model to predict the magnetic phase diagram for larger values of $\mathbf{H}\parallel\mathbf{c}$.  The results are shown in Fig.~\ref{fig:phase_diagram}.  A low temperature metamagnetic transition from FIM-c to a VP-canted structure is predicted at $\mu_0H=23$~T, similar to experimental observations \cite{Suga_2006}.  This transition is driven by a flop of the Er spins into the plane perpendicular to $\mathbf{H}$ (Er-flop). Another metamagnetic transition occurs at $58$~T into the FF phase through which the Er spins flip to be parallel to $\mathbf{H}$  (Er-flip). The first-order character of both transitions are dictated by the large $B_4^0$ Er CEF term which creates an energy barrier between uniaxial ($\theta_{\text{Er}}=0$\degree,$180$\degree) and planar ($\theta_{\text{Er}}=90$\degree) configurations of the Er spins.

    Like the disappearance of the low-field FOMP above $T_{\text{spiral}}$, weakening of the Er anisotropy by thermal fluctuations leads to the disappearance of the VP-canted phase. The transition from FIM-c to FF also becomes continuous at higher temperature. Notably, the ordered Er moment is completely quenched at the FIM-c to FF crossover.  Here, the exchange and Zeeman energies for the Er spins exactly cancel at a crossover field of $\approx12\mathcal{J}^{\text{ME}}(g_J-1)/g_J\mu_B=47$~T. 
	
    Our experimental results and mean-field analysis have revealed the temperature and magnetic field responses of the magnetic order of ErMn$_6$Sn$_6$. Crucial to the compound's magnetic tunability is competition between the various interlayer interactions and single-ion magnetic-anisotropy energies. At zero-field, the increase in thermal spin fluctuations with increasing temperature decreases $\mu_{\text{Er}}$ and results in a weakening of $\mathcal{J}^{\text{ME}}$ and of the uniaxial Er anisotropy.  This leads to the first-order transition from the FIM-ab to a distorted-triple-spiral phase at $T_{\text{spiral}}$. For a nonzero field, weakening of $\mu_{\text{Er}}$ by thermal fluctuations is responsible for the increase and eventual disappearance of the low-field FOMP above $T_{\text{spiral}}$ as well as the elimination of the field-induced VP-canted phase. Remarkably, our calculations also reveal that at high temperatures $\mu_{\text{Er}}$ is effectively quenched at the crossover between FIM-c and FF order due to an exact cancellation of the exchange and Zeeman energies for the Er spins. 

     We have also investigated whether the observed splitting of the lineshapes of the magnetic-Bragg peaks in the triple-spiral phase is explained by the in-plane magnetic anisotropy. This splitting is also observed for Y$166$'s double-spiral order but with a different temperature dependence, begging the question of whether the Er anisotropy explains the differences we observe for Er$166$. Interestingly, recent work on GdV$_6$Sn$_6$ has found evidence for incommensurate amplitude-modulated magnetic order which produces diffraction signatures reminiscent of those seen for the double spiral of Y$166$ \cite{Porter_2023}. The amplitude-modulated order is associated with an RKKY mechanism within the Gd triangular sublattice, however, which is quite different than the Mn-Mn and Mn-Er interlayer exchange we discuss for Er$166$.
     
     We conclude that an unknown distortion of the triple-spiral ordering of Er$166$ exists that is intrinsic to the material and our mean-field analysis verifies that the single-ion anisotropy of the Er, induced by CEF splitting, cannot be responsible in all instances for the distortions. We suggest that a non-sinusoidal layer-to-layer variation of the main spiral and a long-period modulation (beating) exist that occur within each magnetic domain. Similar distortions are often caused by the competition between isotropic-exchange interactions that favor uniform spiral phases and magnetic anisotropy that favors collinear or bunched-spiral distortions \cite{Ramazanoglu_2011,Ratcliff_2016,Chaix_2016}.  Neither spin bunching resulting from anisotropy effects nor amplitude-modulated order, however, is expected to produce even-harmonic satellite diffraction peaks \cite{Chaix_2016} and would not explain the $2\tau$ satellites that we observe. 

   The general understanding of the interactions controlling the low-field spin-reorientation transitions, higher field spin-flop transitions, and the development of conical-spiral phases in $R166$ kagome metals that we have presented gives a roadmap towards studying the behavior of Dirac fermions with control of the Mn spin orientation and real-space spin chirality. In the FIM phases, the low critical field for the FOMP can be a viable source for controlling the Chern gap, which is maximized in the FIM-c phase. For example, magneto-optical transitions between valence and conduction bands of massive Dirac fermions can be switched using small applied fields \cite{Phuphachong_2017,Tokura_2019, Cheskis_2020}.  Triple-spiral phases in Er$166$ with tunable periodicity also are attractive for studying the role of vector spin chirality in transport and optical properties \cite{Nagaosa_2013,Smejkal_2018,Kurumaji_2019}. Indeed, the Hamiltonian parameters listed in Table~\ref{tbl:params} are largely consistent and scalable across the $R$Mn$_6$Sn$_6$ series, with the details of the $R$ magnetic anisotropy and effective magnetic coupling between the $R$ and Mn ions being responsible for determining the magnetic ground state \cite{Lefevre_2002, Clatterbuck_1999}. This latter point is born out by the results of our in-depth study and allows for further predictions of novel magnetic and topological phases across the entire $R166$ series.
    	
	\section{Methods}
	\subsection{Single-Crystal Neutron Diffraction}
	Single-crystals of ErMn$_6$Sn$_6$ (Er166) were grown from excess Sn flux as described previously \cite{Riberolles_2022}.  The samples were determined to be single phase by x-ray diffraction.  Magnetization $M$ measurements were made on a Quantum Design, Inc., Magnetic Property Measurement System down to a temperature of $T=1.8$~K and in magnetic fields up to $\mu_0H=7$~T. For measurements of the $\mathbf{H}\parallel\mathbf{c}$ orientation, the plate-like samples were glued to a plastic disc and held inside a plastic drinking straw. Prior to measuring with the sample, the bare disc was measured for background subtractions.
	
	Single-crystal neutron diffraction measurements were made on the Fixed-Incident-Energy Triple-Axis Spectrometer at the High-Flux Iostope Reactor, Oak Ridge National Laboratory.  Neutrons with a wavelength of $\lambda=2.377$~\AA\ were selected by a double-bounce pyrolitic-graphite (PG) monochromator system and a PG analyzer using the PG $(0,0,2)$ Bragg reflection.  S\"oller slit collimators with collimations of $40^{\prime}$-$40^{\prime}$-$40^{\prime}$-$80^{\prime}$ were placed before the monochromator, between the sample and monochromator, between the sample and analyzer, and between the analyzer and detector, respectively.  PG filters placed after each monochromator were used to reduce contamination by higher-order wavelengths. The sample was mounted on an Al sample holder and cooled in a He closed-cycle refrigerator while either immersed in He exchange gas ($T<300$~K) or in vacuum. Two different samples were studied with masses of  $109.0(1)$~mg and $282.6(1)$~mg over two successive experiments. Both samples were aligned with the $(h,0,l)$ reciprocal-lattice plane set in the scattering plane of the instrument.  Diagrams of the chemical structure were made using \textsc{vesta} \cite{Momma_2011}.
	\subsection{Powder Neutron Diffraction}
	Powder neutron diffraction measurements were made on the time-of-flight diffractometer POWGEN at the Spallation Neutron Source, Oak Ridge National Laboratory. $6.874$~g of Er$166$ powder was loaded into a $6$~mm diameter single-wall vanadium can. The powderized sample was obtained by grinding several single crystals. The sample was cooled using a He closed-cycle refrigerator and the automatic sample changer (PAC) was used.  Data were collected using the high-resolution setting.
	\subsection{Single-Crystal X-Ray Diffraction}
	High-resolution single-crystal x-ray diffraction measurements were performed at Ames National Laboratory using a four-circle diffractometer with Cu $K_{\alpha1}$ radiation from a rotating-anode source and a Ge $(1,1, 1)$ monochromator. The sample was attached to a flat Cu mount which was thermally anchored to the cold head of a He closed-cycle refrigerator. Be domes were used as vacuum shrouds and heat shields. A small amount of He exchange gas facilitated thermal equilibrium.

\section{Data availability}	
The datasets used or analyzed during the current study are  available from the corresponding authors on reasonable request.

\section{Code availability}	
The code used for this study is not publicly available but may be made available to qualified researchers on reasonable request from the corresponding author.

\section{Acknowledgments}	
		SXMR's, TJS's, TH's, PCC's, RJM's, and BGU's work was supported by the Center for the Advancement of Topological Semimetals (CATS), an Energy Frontier Research Center funded by the  U.S.\ Department of Energy (DOE) Office of Science (SC), Office of Basic Energy Sciences (BES), through the Ames National Laboratory. AS's, JMW's, and SLB's work at the Ames National Laboratory is supported by the U.S.\ DOE SC, BES, Division of Materials Sciences and Engineering.  Ames National Laboratory is operated for the U.S.\ DOE by Iowa State University under Contract No.\ DE-AC02-07CH11358. A portion of this research used resources at the High-Flux Isotope Reactor and Spallation Neutron Source, which are U.S.\ DOE SC User Facilities operated by Oak Ridge National Laboratory. This research used resources at the Missouri University Research Reactor (MURR).

\section{Author contributions}
SXMR, TH, WT, QZ, DLA, LDS, RJM, and BGU performed the neutron scattering experiments and analyzed the results. TJS, SLB, and PCC synthesized the samples and performed magnetization measurements. JMW and AS performed single-crystal x-ray diffraction measurements. RJM performed the mean-field analysis. SXMR, RJM, and BGU wrote the manuscript with input from all of the authors.		

\section{Competing interests}
All authors declare no financial or non-financial competing interests.

	

\newpage
    \begin{table*}[h]
	\caption {\textbf{Heisenberg and crystal field parameters for ErMn$_\textbf{6}$Sn$_\textbf{6}$}.  The listed values are given in units of meV and their estimations are described in Supplementary~Section~1.4.}
			\renewcommand\arraystretch{1.25}
	\centering
	\begin{tabular}{ c | c | c | c | c | c | c | c | c | c  }
		\toprule
		$B_2^0$  	& $B_4^0$ & $B_6^0$ 	& $B_6^6$ 	& $K^{\text{M}} $ & $\mathcal{J}_0^{MM}$	& $\mathcal{J}_1^{MM}$ 	& $\mathcal{J}_2^{MM}$ 	& $\mathcal{J}_3^{MM}$ & $\mathcal{J}^{\text{ME}}$	 	\\
		\midrule				
		$0.012$		& $-3.69\times10^{-4}$ & 0				& $1.47\times10^{-5}$		& $0.17$	& $-28.8$		& $-4.4$				&	$-19.2$			& $2.3$					& $1.35$	\\
		\botrule
	\end{tabular}
	\label{tbl:params}
	\end{table*}
		\clearpage

		\clearpage

		\clearpage

		\clearpage

		\clearpage

		\clearpage
	
	\clearpage

		\clearpage
	
	\clearpage
	
	\makeatletter
	
	\renewcommand{\figurename}{Supplementary Figure}
	\renewcommand{\tablename}{Supplementary Table}

	\section{Supplementary Discussion}
	\subsection{Additional Single-Crystal Neutron Diffraction Data} \label{Sec:SXTAL_Diff_SI}
	
	\begin{figure}[]
		\centering
		\includegraphics[width=0.67\linewidth]{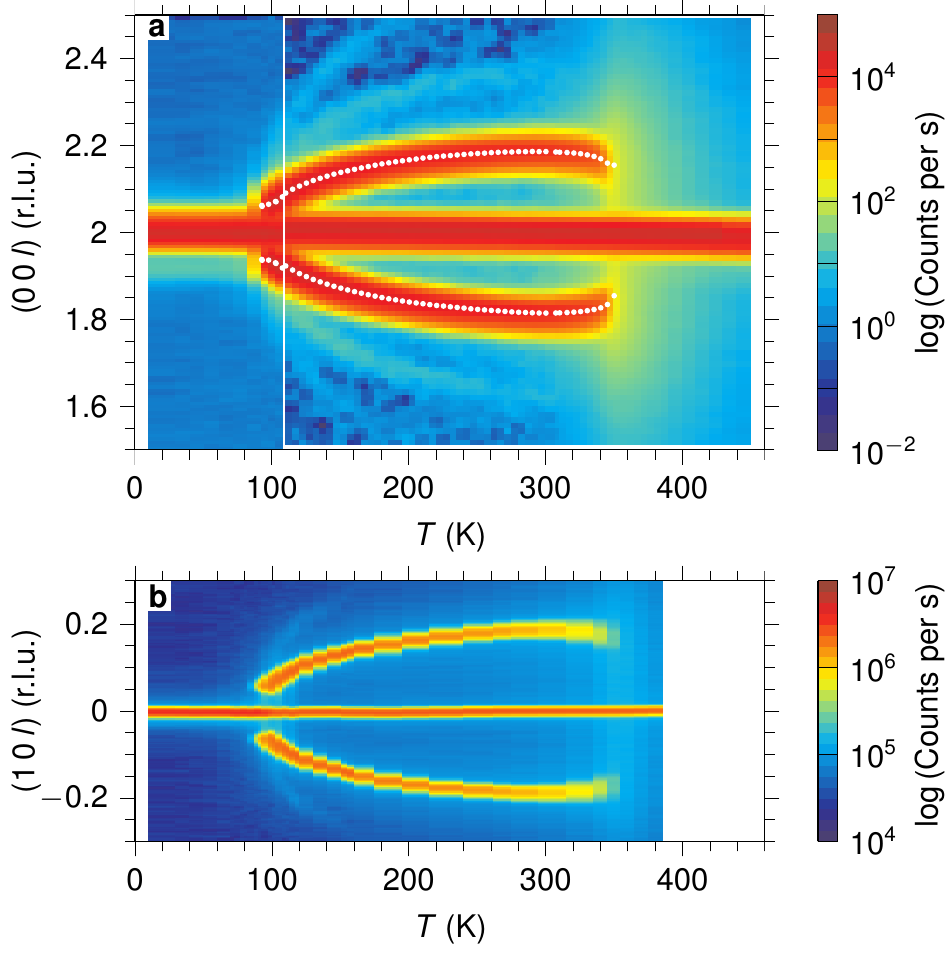}
		\centering
		\caption{  \label{Fig:Cactus_002_100_SI} Temperature evolution of the neutron-diffraction intensity for ErMn$_6$Sn$_6$ from single-crystal neutron diffraction measurements along the \textbf{a} $(0,0,l)$  and  \textbf{b} $(1,0,l)$ reciprocal-lattice directions. The $(1,0,l)$ data were measured during one experiment, whereas the $(0,0,l)$ data were taken over two experiments.  }
	\end{figure}
	
	Supplementary Figure~\ref{Fig:Cactus_002_100_SI} displays the temperature evolution of the magnetic diffraction along $(0,0,l)$ and $(1,0,l)$ to complement Fig.~2 of the main text. The nuclear component of the intensity is significantly weaker for $(1,0,0)$ than for $(0,0, 2)$, and the transition at $T_{\text{spiral}}$ is more clearly seen in the $(1,0,l)$ data.
	
	\subsection{Neutron Powder Diffraction} \label{Sec:Pow_Diff_SI}
	
	\begin{figure}[]
		\centering
		\includegraphics[width=0.67\linewidth]{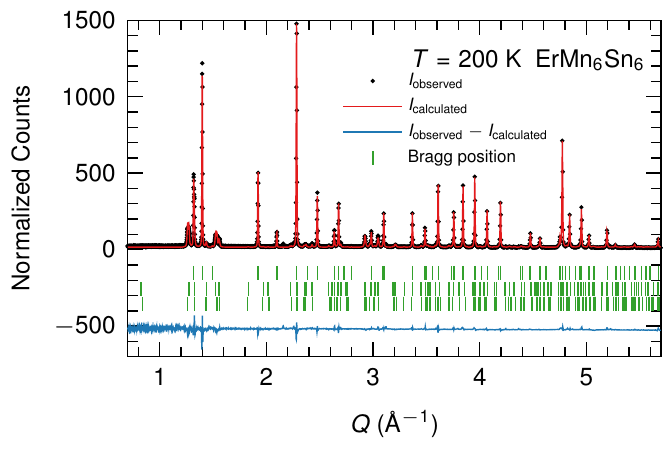}
		\centering
		\caption{Neutron powder diffraction data for ErMn$_6$Sn$_6$ collected at $T=200$~K. The red curve through the data points  shows the fit from a Rietveld refinement using triple-spiral magnetic order and the bottom blue curve shows the difference between the data and fit. Vertical tick marks indicate Bragg peak positions.  The top set of ticks is for the structural-Bragg peaks.  The bottom two sets are for magnetic-Bragg peaks. \label{Fig:pw_refinement_SI}  }
	\end{figure}
	
	\begin{table*}
		\caption {Parameters from Rietveld refinements of neutron powder diffraction data for  ErMn$_6$Sn$_6$ taken at $T=200$~K.  The superscript N indicates a parameter for the structural phase whereas the $+$ and $-$ superscripts indicate parameters for the two magnetic phases described in the text.  The avg superscript indicates the average value for the two magnetic phases. Ordered magnetic moments $\mu_{\text{Er}}$ and $\mu_{\text{Mn}}$ for Er and Mn, respectively, are given in units of $\mu_B$, $\delta$ is given in degrees, and $\tau$ is given in reciprocal-lattice units of $2\pi/c$.}
		
		\centering
		\begin{tabular}{ c | c | c | c | c | c | c |c}
			\hline\hline
			$\mu^{\text{-}}_{\text{Er}}$& $\mu^{\text{-}}_{\text{Mn}}$	& $\delta^{\text{-}}$ &$\tau^{\text{-}}$& $\mu^{\text{+}}_{\text{Er}}$	& $\mu^{\text{+}}_{\text{Mn}}$	& $\delta^{\text{+}}$ &$\tau^{\text{+}}$\\
			\hline		
			$4.8(2)$&$1.8(1)$&$14.3(2)$&$0.1789(4)$&$3.1(2)$&$2.2(1)$&$13.7(2)$&$0.1963(4)$\\			
			\hline\hline
		\end{tabular}
		\begin{tabular}{ c | c | c | c | c | c | c }
			\hline\hline
			$\mu^{\text{avg}}_{\text{Er}}$	& $\mu^{\text{avg}}_{\text{Mn}}$	& $\delta^{\text{avg}}$ & $\tau^{\text{avg}}$ &$R^{\text{N}}_{\text{Bragg}}$ &$R^{\text{-}}_{\text{Bragg}}$ &$R^{\text{+}}_{\text{Bragg}}$\\
			\hline		
			$3.9(3)$ &$2.0(1)$ & $14.0(2)$ & $0.1876(6)$ & $5.36$ & $16.4$ & $12.3$\\			
			\hline\hline
		\end{tabular}
		\label{tbl:pow_refine}
	\end{table*}
	
	Supplementary Figure~\ref{Fig:pw_refinement_SI} shows the neutron powder diffraction pattern for $T=200$~K. Rietveld refinements were made using one phase to describe the chemical structure and two phases to describe the triple-spiral magnetic structure.  Two magnetic phases were used because, as described in the main text, the lineshapes of the magnetic-Bragg peaks are split.  For $200$~K, two gaussian peaks and a background can be used to describe the lineshape. The most straightforward way to model the data for the Rietveld refinements is to assume two equally populated domains, each corresponding to one of the slightly different values of $\bm{\tau}=(0,0,\tau^{\pm})$. Thus, each magnetic phase corresponds to either $\tau^+$ or $\tau^-$ and we refined the values of the ordered magnetic moments for Er ($\mu_{\text{Er}}$) and Mn ($\mu_{\text{Mn}}$), $\tau$, and the angle $(\Phi-\delta)$ for each phase separately. The results are given in Table~\ref{tbl:pow_refine}. The $R_{\text{Bragg}}$ values indicate acceptable agreement between the triple-spiral model and the data. Refinements made allowing $\bm{\mu_{\text{Er}}}$ and $\bm{\mu_{\text{Mn}}}$ to have components along $\mathbf{c}$ did not yield sensible results.
	
	\subsection{Magnetic-Bragg Peak Lineshape Splitting}
	\begin{figure}[]
		\centering
		\includegraphics[width=0.67\linewidth]{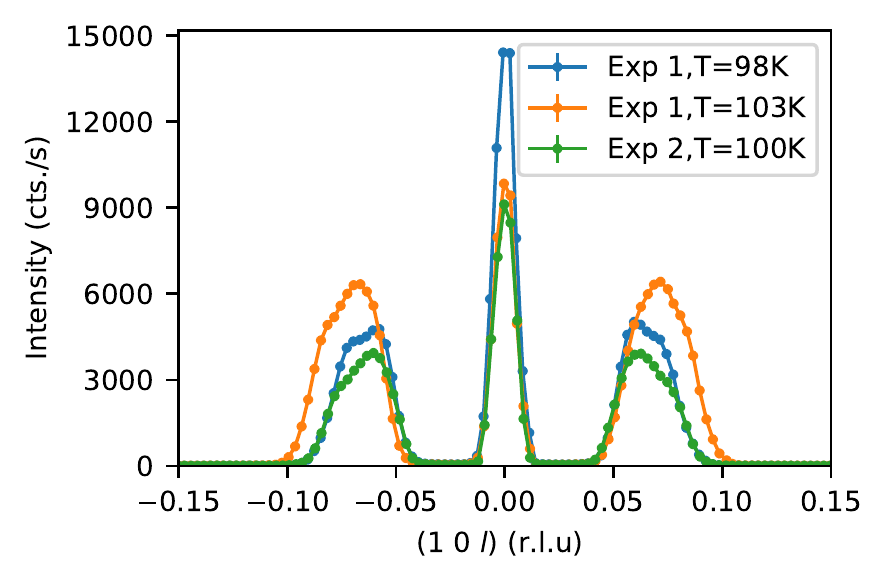}
		\centering
		\caption{ErMn$_6$Sn$_6$ single-crystal neutron diffraction cuts across $(1,0,l)$ in the vicinity of $T_{\text{spiral}}$ from our two experiments. Both samples reveal similar splittings of lineshapes the satellite magnetic-Bragg peaks.  \label{Fig:spliting_Exp1_2_SI}  }
	\end{figure}
	
	Supplementary Figure~\ref{Fig:spliting_Exp1_2_SI} compares $(1,0,l)$ cuts collected in the vicinity of $T_{\text{spiral}}$ during neutron diffraction experiments for two different single-crystal samples of ErMn$_6$Sn$_6$. Fits of the primary-satellites' lineshapes from our first experiment using gaussian-peak components return a difference between the centers of the gaussian components of $\Delta\tau_1=0.015$ at $T=98$~K and $0.017$ at $103$~K. $\Delta\tau_1= 0.016$ is obtained from our second experiment for $100$~K. The close agreement of $\Delta\tau_1$ for two different samples suggests that the lineshape splitting is intrinsic to the compound, and intrinsic to a single magnetic domain, rather than arising from different domains hosting spirals with slightly different periodicity.   
	
	\subsection{Estimation of Hamiltonian parameters} \label{Sec:Ham_Params_SI}
	{\it Exchange parameters}.  Fits to the spin-wave dispersions from previous inelastic-neutron-scattering (INS) data for TbMn$_6$Sn$_6$ (Tb$166$) provide a starting point for estimating the various Mn-Mn and Mn-Er isotropic exchange constants defined in the main text in Eq.~(1).  The values for Tb$166$, in meV,  are; $\mathcal{J}_0^{\text{MM}}=-28.8$, $\mathcal{J}_1^{\text{MM}}=-4.4$, $\mathcal{J}_2^{\text{MM}}=-19.2$, $\mathcal{J}_3^{\text{MM}}=1.8$, and $\mathcal{J}^{\text{MT}}=1.8$~meV \cite{Riberolles_2022,Riberolles_2023}.  Here, $\mathcal{J}_2^{\text{MM}}$ is the dominant ferromagnetic (FM) interlayer coupling. The stability of spiral or ferrimagnetic (FIM) states is then dependent on the balance of the competing $\mathcal{J}_1^{\text{MM}}$ and $\mathcal{J}_3^{\text{MM}}$ interactions, which have opposite signs. We discuss this competition in terms of the ratios $x = \mathcal{J}_1^{\text{MM}}/\mathcal{J}_2^{\text{MM}}>0$ and $y = \mathcal{J}_3^{\text{MM}}/\mathcal{J}_2^{\text{MM}}<0$.  The Tb$166$ value of $x = 0.23$ is consistent with estimates based on neutron diffraction and magnetization data for Y$166$, so we retain this value.  The Tb$166$ value of $y=\left|-0.09\right|$ is smaller than the value of $y=\left|-0.12\right|$ reported for Y$166$.  The smaller of the  two $y$ values is very close to the stability range of the FIM phase and therefore can only stabilize long-period spiral order.  This is not consistent with experimental data in the spiral phase for Er$166$, so we set $y$ to the value reported for Y$166$, or  $\mathcal{J}^{\text{MM}}_3=2.3$~meV.
	
	The Mn-$R$ interaction, $\mathcal{J}^{\text{M}R}$, is known to be antiferromagnetic (AFM) for the heavy rare-earths and both INS data \cite{Riberolles_2022,Riberolles_2023} and density-functional-theory (DFT) calculations \cite{Lee_2023} indicate that $\mathcal{J}^{\text{M}R}$ decreases in magnitude from $R=\text{Gd}\rightarrow\text{Tm}$. We estimate the value of $\mathcal{J}^{\text{ME}}=1.35$~meV using INS data for Er$166$ and magnetization data, as described below. This value is lower than reported values for Gd$166$ ($\mathcal{J}^{\text{MG}}=2$~meV) and Tb$166$ ($\mathcal{J}^{\text{MT}}=1.8$~meV) which is consistent with the expected trend.
	
	{\it Mn easy-plane anisotropy}. Mn has an easy-plane (planar) anisotropy with $K^{\text{M}}>0$.  The magnitude of $K^{\text{M}}$ in $R166$ compounds has been estimated to be in a range from $0.2$ to $0.5$~meV from a variety of experimental approaches; ($1$) fitting the spin gap of Tb$166$ results in $K^{\text{M}}=0.44$~meV \cite{Riberolles_2023}; ($2$) reported from analysis of the magnetization data as $0.47$~meV for Tb$166$ \cite{Jones_2022} ($3$) reported to be $0.23$~meV from the $\mathbf{H}\parallel\mathbf{c}$ saturation magnetization field of Gd$166$ \cite{Rosenfeld_2008}; ($4$) reported as $0.2$~meV for Y$166$ \cite{Ghimire_2020}.
	
	For Er$166$, the best estimate of $K^{\text{M}}$ is obtained from the high-temperature magnetization data with $\mathbf{H}\parallel\mathbf{c}$, as shown in Fig.~7a of the main text.  At temperatures where the Er MAE is quenched ($T\sim 200$~K), the critical field for ferrimagnetic alignment of Mn and Er moments is approximately $\mu_0H_{\text{c}}=5$~T.  If this critical field is determined solely by the MAE of the Mn ion, we estimate that $K^{\text{M}}\approx\mu_0H_{\text{c}}/12M = 0.17$~meV, where $M=7~\mu_B$ is the onset of the magnetization plateau at $200$~K.  This value of $K^{\text{M}}$ is on the low end of the range of the other $R166$ compounds and so we fix $K^{\text{M}}=0.17$ meV.
	
	{\it Er crystalline-electric-field parameters}.  The crystalline-electric-field (CEF) parameters $B_l^m$ for different rare-earth ions scale according to the formula
	\begin{equation}
		B_l^m=\langle r^l \rangle \theta_l A_l^m
		\label{eqn:Stevens}
	\end{equation}
	where $\theta_l$ is the Steven's factor and $\langle r^l \rangle$ is the average $l^{\text{th}}$ order radial moment, both of which depend on the $R$ ion.  The parameter $A_l^m$ is intrinsic to the crystalline potential and is expected to vary slowly across the $R166$ series.
	
	From INS studies of Tb$166$, we are able to determine approximate values of $A_2^0 = (4.1~\text {meV} )a_0^{-2}$, $A_4^0 = -(7.0~\text{meV}) a_0^{-4}$, and $A_6^0 \approx 0$ (where $a_0$ is the Bohr radius).   Assuming the transferability of the CEF potential for the hexagonal $R166$s and using Eq.~\eqref{eqn:Stevens}, we obtain good starting values for the CEF parameters for the Er ion; $B_2^0=0.0075$~meV, $B_4^0=-0.00041$~meV and $B_6^0=0$.  These values are reasonably consistent with DFT studies of the $R$Mn$_6$Sn$_6$ series \cite{Lee_2023} and slight refinement of these values are described below.  For Er$166$, magnetization measurements indicate an easy-axis along the $(1,1,0)$ direction in the hexagonal unit cell (see Ref.~\cite{Suga_2006}) implying that the in-plane anisotropy term $B_6^6$ is positive ($B_6^6>0$). 
	
\begin{figure}
		\includegraphics[width=0.67\linewidth]{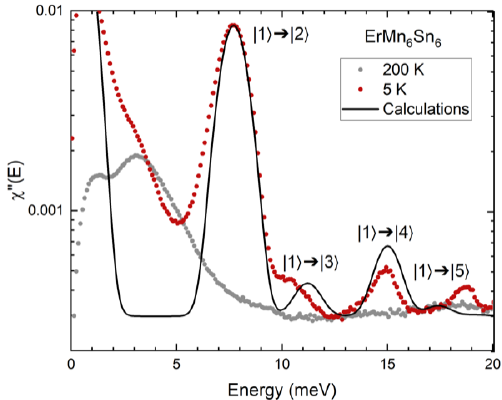}
		\centering
		\caption{Comparison of inelastic-neutron-scattering (INS) data at $T=5$~K and $200$~K to calculations of the trace of the imaginary part of the local-ion magnetic susceptibility (black line) using the parameters determined in Supplementary Figure~\ref{fig:global_SI}.  INS data were measured on the ARCS spectrometer at the Spallation Neutron Source using an incident energy of $30$~meV.  Data are summed over $h=0.25$ to $0.75$, $k=-0.25$ to $0.25$, and $l=-5$ to $5$ for the hexagonal reciprocal lattice. }
		\label{fig:CEF_INS_SI}
	\end{figure}

	{\it Global search}.  Next we describe estimates of $B_l^m$ and $\mathcal{J}^{\text{ME}}$ for Er$166$ from comparison of magnetization and INS data.  In this search, we fix the other parameters of the model (i.e.\ $\mathcal{J}_i^{\text{MM}}$ and $K^{\text{M}}$). Supplementary Figure~\ref{fig:CEF_INS_SI} shows data from INS measurements of Er$166$ where we observe four CEF excitations out of the ground state at $T=5$~K.  These excitations are not visible at $200$~K due to thermal depopulation of the ground state. The energies of the first four transitions were obtained from gaussian fits to the peaks and are listed in Table~\ref{tbl:obs}.
	
	As shown in Fig.~7a of the main text, an applied magnetic field along the hard $(0,0,1)$ direction will generate a first-order magnetization process (FOMP) where the ground-state FIM structure jumps from spins ordered with their orientations in the $ab$-plane (planar) to their orientations lying along the $c$-axis (uniaxial).  The small value of the critical FOMP field [$\mu_0 H_{\text{c}} = 0.7(2)$~T] indicates that the planar and uniaxial configurations are nearly degenerate with a free energy difference of $\Delta\mathcal{F}=\mathcal{F}_{\text{c}}-\mathcal{F}_{\text{ab}} \approx M\mu_0 H_{\text{c}} = 0.16(5)$~meV, where $M=4~\mu_B/\text{fu}$ is the net magnetization per formula unit in the FIM ground state. 
	
\begin{figure}
		\includegraphics[width=0.67\linewidth]{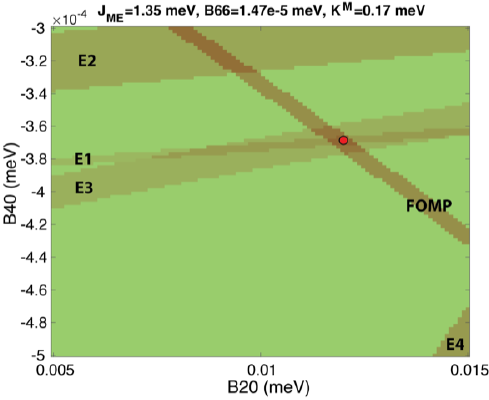}
		\centering
		\caption{Global search for model parameters as a function of $B_2^0$ and $B_4^0$ for fixed values of $\mathcal{J}^{\text{ME}}=1.35$~meV, $B_6^6=1.47\times10^{-5}$~meV, and $K^{\text{M}}=0.17$~meV, and all other exchange constants fixed as described in the text.  Striped regions correspond to observables listed in Table~\ref{tbl:obs}.  The red dot shows a choice of parameters for $B_2^0$ and $B_4^0$ that best represents the first-order magnetization process (FOMP) and the $E1$, and $E_3$  crystalline-electric-field transitions.}
		\label{fig:global_SI}
	\end{figure}

	We can use a combination of INS and magnetization data to refine our estimates of these parameters.  Supplementary Figure~\ref{fig:global_SI} shows a representative slice of a search in the $4$-D space of $B_l^m$ and $\mathcal{J}^{\text{ME}}$ parameters. The indicated regions for the different observables correspond to the values in Table~\ref{tbl:obs}.  In Supplementary Figure~\ref{fig:global_SI}, the selected parameters are indicated by the red dot and provide good estimates of $\Delta\mathcal{F}$, $E_1$, and $E_3$, whereas agreement with $E_2$ and $E_4$ is less optimal.  The black curve in Supplementary Figure~\ref{fig:CEF_INS_SI} compares the calculated CEF transitions from this parameter set to the INS data.  The main text demonstrates that our model with these parameters displays reasonable agreement with the observed data and is valuable in interpreting experimental results.
	
	\begin{table*}
		\caption {Energies obtained from magnetization data [$\Delta\mathcal{F}\text{(FOMP)}$] and crystalline-electric-field (CEF) transitions ($Ei$) seen in inelastic-neutron-scattering data.  These energies are used to determine the Hamiltonian parameters for Er$166$. The numbers for the CEF states are general labels starting from the ground state $|1\rangle$.}
		\renewcommand\arraystretch{1.25}
		\centering
		\begin{tabular}{ c | d{-1}  }
			\hline\hline
			Observable & \multicolumn{1}{c}{Energy (meV)} \\
			\hline
			$\Delta\mathcal{F}$  (FOMP)			&0.16(5)\\
			$E1 (|1\rangle \rightarrow |2\rangle)$		&7.65(1)\\
			$E2 (|1\rangle \rightarrow |3\rangle)$		&10.50(15)\\
			$E3 (|1\rangle \rightarrow |4\rangle)$		&14.85(13)\\
			$E4 (|1\rangle \rightarrow |5\rangle)$		&18.6(3)\\
			\hline\hline
		\end{tabular}
		\label{tbl:obs}
	\end{table*}
	
	\subsection{Classical magnetic anisotropy energy}\label{Sec:Class_MAE_SI}
	The classical magnetic anisotropy energy (MAE) for hexagonal Er$166$ is given by
	\begin{align}
		K_{\text{tot}} & =  K_1\sin^2\theta_{\text{Er}} + K_2\sin^4\theta_{\text{Er}} +  K_3\sin^6\theta_{\text{Er}} \\ \nonumber
		& + K_3^\prime\cos(6\varphi_{\text{Er}})\sin^6\theta_{\text{Er}} - 6K^{\text{M}}\sin^2\theta_{\text{Mn}}.
	\end{align}
	where $\theta_i$ and $\varphi_i$ are the polar and azimuthal angles, respectively, describing the orientation of the Er ($i=\text{Er}$) and Mn ($i=\text{Mn}$) magnetic moments.  For Mn, we retain only the first-order MAE term and for Er we assume that $K_3=0$.  The MAE constants for Er are related to the $B_l^m$ parameters, $K_1=-3J^{(2)}B_2^0-40J^{(4)}B_4^0$, $K_2=35J^{(4)}B_4^0$, and $K_3^{\prime}=J^{(6)}B_6^6$, where $J^{(2)}=J(J-1/2)$, $J^{(4)}=J^{(2)}(J-1)(J-3/2)$, and $J^{(6)}=J^{(4)}(J-2)(J-5/2)$. 
	
	We find that $K_{\text{tot}}=0$ for a uniaxial ferrimagnet (FIM-c) with $\theta_{\text{Er}}=\pi$ and $\theta_{\text{Mn}}=0$ and $K_{\text{tot}} = K_1+K_2-K_3^{\prime}-6K^{\text{M}}<0$ for a planar ferrimagnet (FIM-ab) with the in-plane easy-axis defined by $\theta_{\text{Er}}=\theta_{\text{Mn}}=\pi/2$ and $\varphi_{\text{Er}}=\pi/6$.   Using our Er CEF parameters, we obtain $K_1=28.34$, $K_2=-26.45$, $K_3^{\prime}=0.83$, and $6K^{\text{M}}=1.02$ meV.  The classical polar and planar MAE are plotted in Supplementary Figures~\ref{fig:MAE_SI}a and \ref{fig:MAE_SI}b, respectively, and demonstrate the near degeneracy of uniaxial and planar configurations which arise from an almost complete cancellation of Mn and Er MAE contributions. We find that the classical MAE slightly favors the FIM-c phase ($K_{\text{tot}}=0.04$ meV), whereas experimentally FIM-ab is the ground state and we estimate $K_{\text{tot}} \approx -M\mu_0 H_{\text{FOMP}}/2 = -0.08$ meV from the FOMP field, as described above. 
	
	\begin{figure}
		\includegraphics[width=0.6\linewidth]{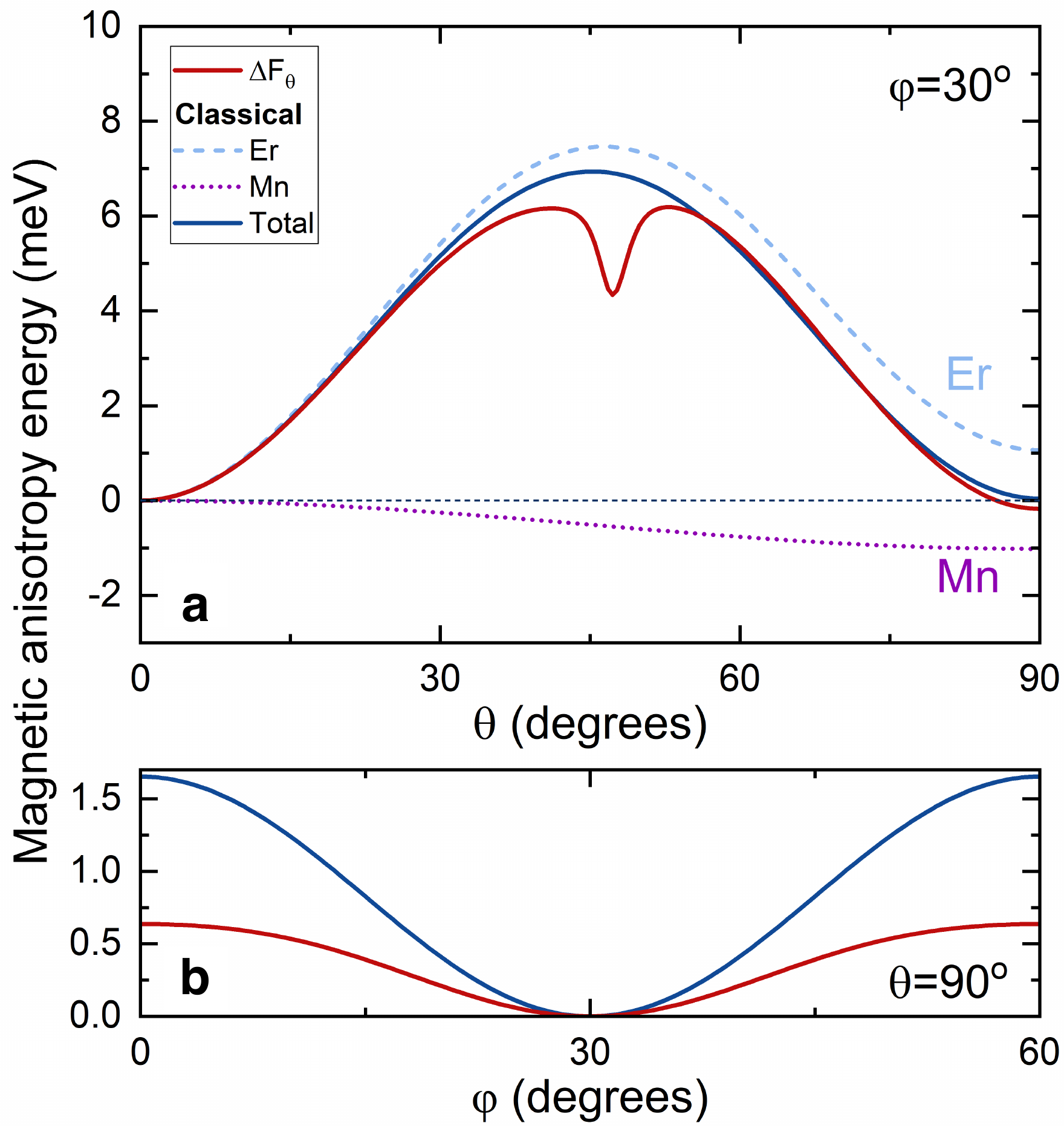}
		\centering
		\caption{Ground state magnetic anisotropy energy (MAE) of ErMn$_6$Sn$_6$ as a function of \textbf{a} the polar angle ($\theta$) and \textbf{b} the planar angle ($\varphi$) in both the classical limit (blue) and from free energy calculations (red). In \textbf{a}, the separate contributions of Er (blue dash) and Mn (purple dot) to the classical MAE are shown.}
		\label{fig:MAE_SI}
	\end{figure}
	
	Evaluation of the MAE using a mean-field approach does establish the FIM-ab state as the ground state. Furthermore, orbital mixing of the CEF eigenstates reduces $\mu_{\text{Er}}$ and leads to differences between the classical and mean-field results.  This is illustrated by the small differences between the classical and mean-field curves shown in Fig.~\ref{fig:MAE_SI}a for a progressively tilted collinear-FIM structure with $\theta_{\text{Mn}}=\theta$ and $\theta_{\text{Er}}=\pi-\theta$ at $T=0$~K.  Similarly, Supplementary Figure~\ref{fig:MAE_SI}b compares the classical and free-energy results for the Er six-fold planar MAE, demonstrating that the easy axis is rotated $30$\degree\ away from $\mathbf{a}$. Note that the CEF level mixing captured in the mean-field approach leads to a sizable reduction of the planar MAE due to its dependence on the sixth-order matrix elements of the angular-momentum operators.
	
	\subsection{Mean-field description of the free energy}\label{Sec:Mean_Field_SI}
	The typical mean-field decomposition of the exchange Hamiltonian Eq.~(1) of the main text is given by, 
	\begin{align}
		\mathcal{H}_{\text{ex}}^{\text{MF}} = &\mathcal{J}^{\text{ME}} \sum_{\langle i<j \rangle} \left(\langle\mathbf{s}_i \rangle \cdot \mathbf{S}_j  + \mathbf{s}_i \cdot \langle \mathbf{S}_j\rangle - \langle \mathbf{s}_i \rangle \cdot \langle \mathbf{S}_j\rangle\right)   \nonumber \\
		+& \sum_{i,j} \mathcal{J}^{\text{MM}}_{ij} \left(\langle \mathbf{s}_i\rangle \cdot \mathbf{s}_j + \mathbf{s}_i \cdot \langle \mathbf{s}_j \rangle -\langle \mathbf{s}_i \rangle \cdot \langle \mathbf{s}_j \rangle\right) .
		\label{Hex_MF}
	\end{align}
	
	We use these terms to generate local mean-field Hamiltonians for Er and Mn 
	\begin{align}
		\mathcal{H}_{\text{Er}}^{\text{MF}}&= \mathcal{H}_{\text{Er}} - {\bf B}_{\text{Er}} \cdot \mathbf{S} - g_J\mu_{\text B}{\bf J} \cdot \mu_0{\bf H} 
		\label{HEr_MF} 
		\intertext{and}
		\mathcal{H}_{\text{Mn}}^{\text{MF}}&= \mathcal{H}_{\text{Mn}} - {\bf B}_{\text{Mn}} \cdot \mathbf{s} - g\mu_{\text B}{\bf s} \cdot \mu_0{\bf H}\, ,
		\label{HMn_MF}
	\end{align}
	where ${\bf B}_{\text{Mn}}= -2\mathcal{J}^{\text{ME}} \langle\mathbf{S} \rangle - \sum_i\gamma_i \mathcal{J}^{\text{MM}}_{i} \langle \mathbf{s}_i \rangle$ and ${\bf B}_{\text {Er}}=-12\mathcal{J}^{\text{ME}} \langle\mathbf{s}\rangle$  are the self-consistently determined molecular fields acting on Mn and Er, respectively.   $\gamma_i$ is the coordination number for $\mathcal{J}^{\text{MM}}_{i}$. 
	
	The self-consistent solution to the mean-field Hamiltonian is obtained by minimizing the free energy
	\begin{align}
		\mathcal{F}(T,H_z) &=-k_BT\ln Z_{\text{Er}} - 6k_BT\ln Z_{\text{Mn}}  \\ \nonumber
		& - 12\mathcal{J}^{\text{ME}} \langle\mathbf{s}\rangle \cdot \langle \mathbf{S} \rangle - 3\sum_i \gamma_i \mathcal{J}^{\text{MM}}_{i} \langle \mathbf{s}_i \rangle \cdot \langle \mathbf{s} \rangle\ ,
		\label{Free}
	\end{align}
	where $Z_{\text{Er}}$ and $Z_{\text{Mn}}$ are the partition functions for Er and Mn obtained from  Eqs.~\eqref{HEr_MF} and \eqref{HMn_MF}, respectively. 
	
	To simplify our analysis, we consider only uniaxial fields $\mathbf{H}\parallel\mathbf{c}$ (i.e.\ $H_z$) and make an approximation where the planar Er magnetic-anisotropy energy (MAE) given by $K_3^{\prime}\sin^6\theta_{\text{Er}}\cos(6\varphi_{\text{Er}})$ (where $K_3^{\prime} \propto B_6^6$) is fixed at its minimum value of $-K_3^{\prime}\sin^6\theta_{\text{Er}}$ for any in-plane angle $\varphi_{\text{Er}}$.  This is equivalent to an easy-plane anisotropy with nonzero $B_6^0$, as the $\theta$ dependence plays a critical role in controlling the phase stability. Furthermore, Fig.~6a in the main text shows that the true easy-plane limit ($K_3^{\prime}\approx0$) is valid at high temperatures ($T>100$~K) due to rapid thermal softening of the planar MAE upon warming. This approximation effectively describes the low-temperature magnetic phases, such as FIM, where all moments lie in a vertical plane containing the planar easy-axis and $c$-axis field directions. 
	
\begin{figure}
		\includegraphics[width=0.5\linewidth]{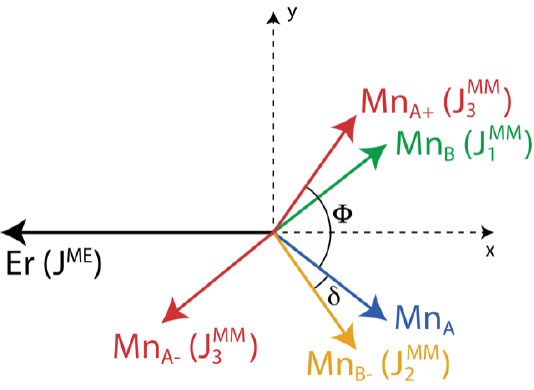}
		\centering
		\caption{Geometry of spins in neighboring layers in the easy-plane triple-spiral state.  We place the Er spin (black) arbitrarily along the $(\bar{1},0,0)$ direction and it's molecular field is determined by the $\mathcal{J}^{\text{ME}}$ coupling to the Mn$_\text{A}$ (blue) and Mn$_\text{B}$ (green) in the layers above and below.  The blue Mn$_\text{A}$ spin experiences a molecular field from the Mn$_\text{A+}$ and Mn$_\text{A-}$ (red), Mn$_\text{B-}$ (yellow), Mn$_\text{B}$ (green) and the Er spin with the exchange parameters indicated in parentheses.}
		\label{fig:geom_SI}
	\end{figure}

	{\it Zero-field case}. For easy-plane configurations [$K_3^{\prime}\cos(6\varphi_{\text{Er}})=const.$]  in zero field, only the relative angles between spins matter. In the specific case of the ideal-triple-spiral structure, the periodicity is defined by the angle $\Phi$ between alike layers in adjacent cells (coupled by $\mathcal{J}^{\text{MM}}_3$) and the angle $\delta$ is between strongly coupled Mn bilayers (coupled by $\mathcal{J}^{\text{MM}}_2$). Supplementary Figure~\ref{fig:geom_SI} (and Fig.~1 of the main text) shows the relative angles of spins in neighboring layers and their coupling constants.  In Supplementary Figure~\ref{fig:geom_SI}, the Er moment is arbitrarily chosen to point along $(\bar{1},0,0)$, bisecting the angle ($\Phi-\delta$) and the Mn moment directions.  The Mn moments are specified by layer designations $\text{A}$, $\text{B}$, $\text{A+}$, $\text{A-}$, and $\text{B-}$ and their orientations are given relative to the Er moment orientation. The spin vectors are 
	
	\begin{align}
		\mathbf{S} &= S(\bar{1},0,0)\,, \\ \nonumber
		\mathbf{s}_\text{A} &= s\left[ \cos \left(\frac{\Phi-\delta}{2}\right) , -\sin \left(\frac{\Phi-\delta}{2}\right),0\right] \,,\\ \nonumber
		\mathbf{s}_\text{B} &= s\left[ \cos \left(\frac{\Phi-\delta}{2}\right) ,  \sin \left(\frac{\Phi-\delta}{2}\right),0\right] \,,\\ \nonumber
		\mathbf{s}_\text{A-} &= s\left[ \cos \left(\frac{3\Phi-\delta}{2}\right) , - \sin \left(\frac{3\Phi-\delta}{2}\right),0\right]\,, \\ \nonumber
		\mathbf{s}_\text{A+} &= s\left[ \cos \left(\frac{\Phi+\delta}{2}\right) ,  \sin \left(\frac{\Phi+\delta}{2}\right), 0\right]\,, \\ \nonumber
		\mathbf{s}_\text{B-} &= s\left[ \cos \left(\frac{\Phi+\delta}{2}\right) ,- \sin \left(\frac{\Phi+\delta}{2}\right), 0\right]  \nonumber \,,
	\end{align}
	and Supplementary Figure~\ref{fig:geom_SI} shows the relative angles of spins in neighboring layers and their coupling constants.
	
	In the zero-field case, expressions for the molecular fields are 
	\begin{align}
		\mathbf{B}_\text{Er} &= -6\mathcal{J}^{\text{ME}} \left(\langle\mathbf{s}_\text{A}\rangle + \langle\mathbf{s}_\text{B}\rangle\right) \,,\\ \nonumber
		& = -12\mathcal{J}^{\text{ME}}\langle{s}\rangle \cos \left(\frac{\Phi-\delta}{2}\right) (1,0,0) \,,
		\intertext{and} 
		\mathbf{B}_\text{Mn} &=  -2\mathcal{J}^{\text{ME}} \langle\mathbf{S}\rangle - 4\mathcal{J}_0^{\text{MM}}\langle\mathbf{s}_\text{A}\rangle - \mathcal{J}_1^{\text{MM}}\langle\mathbf{s}_\text{B}\rangle \\ \nonumber
		& - \mathcal{J}_2^{\text{MM}}\langle\mathbf{s}_\text{B-}\rangle - \mathcal{J}_3^{\text{MM}}\left(\langle\mathbf{s}_\text{A+}\rangle + \langle\mathbf{s}_\text{A-}\rangle\right) \,,\\ \nonumber
		& = 2\mathcal{J}^{\text{ME}}\langle{S}\rangle (1,0,0) \\ \nonumber
		& - \left(4\mathcal{J}_0^{\text{MM}}+2\mathcal{J}_3^{\text{MM}}\cos \Phi\right)\langle s\rangle\\ \nonumber
		&\times\left [ \cos \left(\frac{\Phi-\delta}{2}\right) , - \sin \left(\frac{\Phi-\delta}{2}\right), 0\right] \\ \nonumber
		& - \mathcal{J}_1^{\text{MM}}\langle s\rangle \left[\cos \left(\frac{\Phi-\delta}{2}\right) , \sin \left(\frac{\Phi-\delta}{2}\right), 0\right] \\ \nonumber
		& - \mathcal{J}_2^{\text{MM}}\langle s\rangle \left[\cos \left(\frac{\Phi+\delta}{2}\right) ,-\sin \left(\frac{\Phi+\delta}{2}\right), 0\right]  \nonumber\,.
	\end{align}
	The resulting free energy per unit cell is given by
	\begin{align}
		&\mathcal{F} = -k_BT\ln Z_\text{Er} - 6k_BT\ln Z_\text{Mn}  \\ \nonumber
		& + 12\mathcal{J}^{\text{ME}} \langle s\rangle \langle S\rangle \cos\left(\frac{\Phi-\delta}{2}\right) \\ \nonumber
		& - 3 \langle s \rangle^2\left[4\mathcal{J}_0^{\text{MM}} + \mathcal{J}^{\text{MM}}_1\cos(\Phi-\delta)+\mathcal{J}^{\text{MM}}_2\cos\delta \right. \\ \nonumber
		& \left.+2\mathcal{J}^{\text{MM}}_3\cos\Phi\vphantom{4\mathcal{J}_0^{\text{MM}} }\right] \nonumber \,,
		\label{Free_triple}
	\end{align}
	where the thermally averaged spin magnitudes $\langle s\rangle$ and $\langle S\rangle$ are determined self-consistently.
	
	{\it Vertical-field case}. For a vertical magnetic field along the $\mathbf{z}$ (i.e.\ $\mathbf{c}$) direction, we maintain the double spiral angles $\delta$ and $\Phi$ and now introduce the polar angles of the Er and Mn sublattices $\theta_{\text{Er}}$ and $\theta_{\text{Mn}}$. Similar to the definitions above, we assume that the Er moment is pointed in the $(1,0,1)$ plane  and always negative in the $x-$direction. The spin vectors are \begin{align}
		\mathbf{S} &= S\left[-\sin \theta_{\text{Er}} , 0 , \cos \theta_{\text{Er}}\right] \,,\\ \nonumber
		\mathbf{s}_\text{A} &= s\left[\sin \theta_{\text{Mn}} \cos \left(\frac{\Phi-\delta}{2}\right) , -\sin \theta_{\text{Mn}} \sin \left(\frac{\Phi-\delta}{2}\right), \cos \theta_{\text{Mn}}\right] \,,\\ \nonumber
		\mathbf{s}_\text{B} &= s\left[\sin \theta_{\text{Mn}} \cos \left(\frac{\Phi-\delta}{2}\right) , \sin \theta_{\text{Mn}} \sin \left(\frac{\Phi-\delta}{2}\right), \cos \theta_{\text{Mn}}\right] \,,\\ \nonumber
		\mathbf{s}_\text{A-} &= s\left[\sin \theta_{\text{Mn}} \cos \left(\frac{3\Phi-\delta}{2}\right) , -\sin \theta_{\text{Mn}} \sin \left(\frac{3\Phi-\delta}{2}\right), \cos \theta_{\text{Mn}}\right]\,, \\ \nonumber
		\mathbf{s}_\text{A+} &= s\left[\sin \theta_{\text{Mn}} \cos \left(\frac{\Phi+\delta}{2}\right) , \sin \theta_{\text{Mn}} \sin \left(\frac{\Phi+\delta}{2}\right), \cos \theta_{\text{Mn}}\right] \,,\\ \nonumber
		\mathbf{s}_\text{B-} &= s\left[\sin \theta_{\text{Mn}} \cos \left(\frac{\Phi+\delta}{2}\right) ,-\sin \theta_{\text{Mn}} \sin \left(\frac{\Phi+\delta}{2}\right), \cos \theta_{\text{Mn}}\right] \,. \nonumber
	\end{align}
	
	The corresponding molecular fields are given by similar expressions as above and include an externally applied magnetic field along the $z-$axis.
	
	\begin{align}
		\mathbf{B}_\text{Er} &= -12\mathcal{J}^{\text{ME}}\langle{s}\rangle \left[\sin \theta_{\text{Mn}} \cos \left(\frac{\Phi-\delta}{2}\right) , 0, \cos \theta_{\text{Mn}}\right]  \nonumber \\
		+& \mu_0H_z(0,0,1)\,,\\
		\mathbf{B}_\text{Mn} &= -2\mathcal{J}^{\text{ME}}\langle S\rangle [-\sin \theta_{\text{Er}} , 0,  \cos \theta_{\text{Er}}] \nonumber \\ \nonumber
		- &4\mathcal{J}_0^{\text{MM}}\langle s\rangle \\ \nonumber
		\times&\left[\sin \theta_{\text{Mn}} \cos \left(\frac{\Phi-\delta}{2}\right) , -\sin \theta_{\text{Mn}} \sin \left(\frac{\Phi-\delta}{2}\right), \cos \theta_{\text{Mn}}\right] \\ \nonumber
		-& \mathcal{J}_1^{\text{MM}}\langle s\rangle\\ \nonumber
		\times& \left[\sin \theta_{\text{Mn}} \cos \left(\frac{\Phi-\delta}{2}\right) , \sin \theta_{\text{Mn}} \sin \left(\frac{\Phi-\delta}{2}\right), \cos \theta_{\text{Mn}}\right] \\ \nonumber
		-& \mathcal{J}_2^{\text{MM}}\langle s\rangle \\ \nonumber
		\times&\left[\sin \theta_{\text{Mn}} \cos \left(\frac{\Phi+\delta}{2}\right) ,-\sin \theta_{\text{Mn}} \sin \left(\frac{\Phi+\delta}{2}\right), \cos \theta_{\text{Mn}}\right] \\ \nonumber
		-& 2\mathcal{J}_3^{\text{MM}}\langle s\rangle \left[\sin \theta_{\text{Mn}} \cos \Phi \cos\left(\frac{\Phi-\delta}{2}\right),\right.\ldots \\  \nonumber 
		&\ldots\left. -\sin \theta_{\text{Mn}} \cos \Phi \sin \left(\frac{\Phi-\delta}{2}\right), \cos \theta_{\text{Mn}}\vphantom{\left(\frac{\Phi-\delta}{2}\right)}\right] \\
		+& \mu_0H_z(0,0,1)\,.
	\end{align}
	Using the same notation, the free energy is
	\begin{align}
		\mathcal{F} &= -k_BT\ln Z_\text{Er} - 6k_BT\ln Z_\text{Mn}  \\ \nonumber
		& + 12 \mathcal{J}_0^{\text{MM}} \langle s\rangle^2 + 12\mathcal{J}^{\text{ME}} \langle S\rangle \langle s\rangle\\ \nonumber
		&\times \left[-\sin \theta_{\text{Mn}} \sin \theta_{\text{Er}}  \cos \left(\frac{\Phi-\delta}{2}\right) + \cos \theta_{\text{Mn}} \cos \theta_{\text{Er}}\right] \\ \nonumber
		& + 3  \mathcal{J}_1^{\text{MM}} \langle s\rangle^2 \left[\sin^2\theta_{\text{Mn}} \cos(\Phi-\delta) + \cos^2\theta_{\text{Mn}}\right] \\ \nonumber
		& + 3  \mathcal{J}_2^{\text{MM}} \langle s\rangle^2 \left(\sin^2\theta_{\text{Mn}} \cos\delta + \cos^2\theta_{\text{Mn}}\right) \\ \nonumber
		& + 6  \mathcal{J}_3^{\text{MM}} \langle s\rangle^2 \left(\sin^2\theta_{\text{Mn}} \cos\Phi + \cos^2\theta_{\text{Mn}}\right)\,.
	\end{align}
	
	Supplementary Figure~\ref{fig:mag_high_SI} shows the evolution of the polar angles and total magnetization that result from minimization of the free energy for $T=0$ and $T=125$~K as a function of applied field.  Similar calculations were used to assemble  the phase diagram shown in Fig.~8 of the main text. 
	
	\begin{figure*}
		\includegraphics[width=0.9\linewidth]{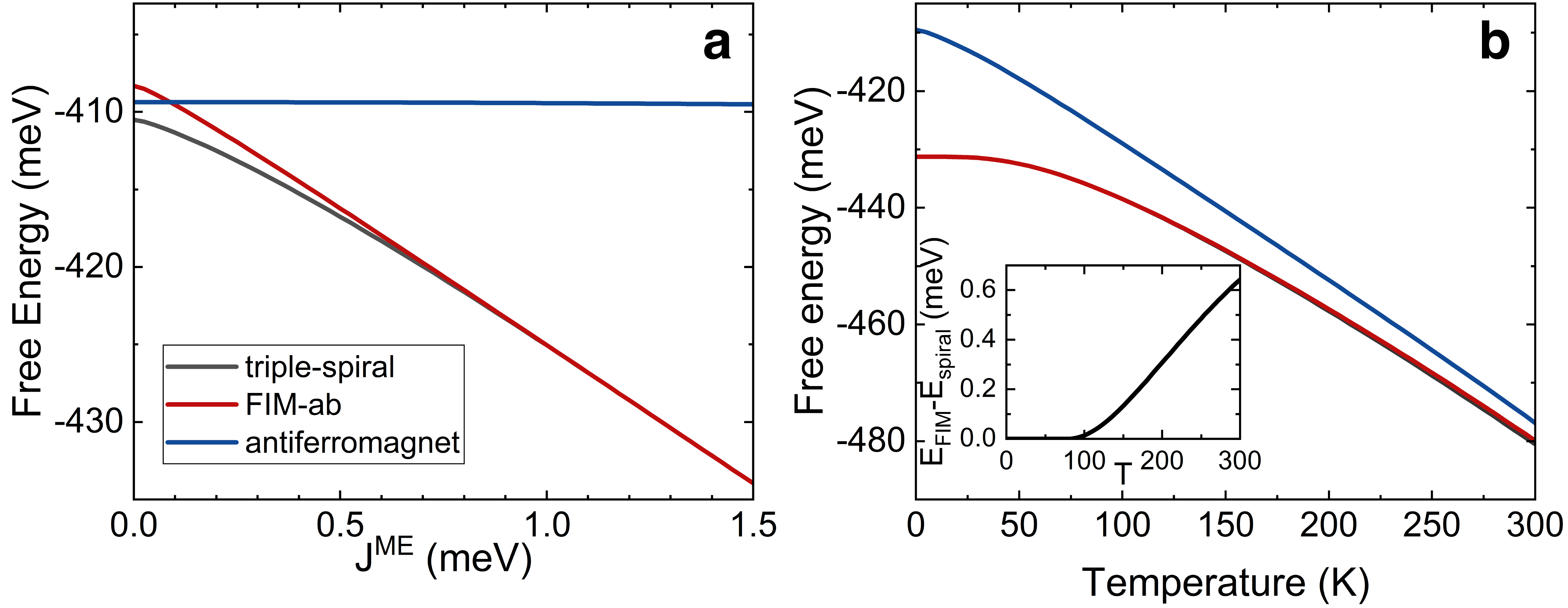}
		\centering
		\caption{ \textbf{Free energies of the zero-field magnetic phases in ErMn$_\textbf{6}$Sn$_\textbf{6}$.} We calculate the phase stability of different planar ($\theta_{\text{Er}}=\theta_{\text{Mn}}=\pi/2$) ferrimagnetic (FIM-ab), antiferromagnetic, and spiral phases using a mean-field analysis of the easy-plane model for ErMn$_6$Sn$_6$.  \textbf{a} Free energy of different planar magnetic ground states at $T=0$~K as a function of $\mathcal{J}^{\text{ME}}$. \textbf{b} The temperature dependence of the phase stability of the planar states with $\mathcal{J}^{\text{ME}}=1.35$~meV.  The inset to \textbf{b} shows the small free energy difference between the FIM-ab and triple-spiral phases.}
		\label{fig:mean_field_xy_SI}
	\end{figure*}
	
	\begin{figure}
		\includegraphics[width=0.5\linewidth]{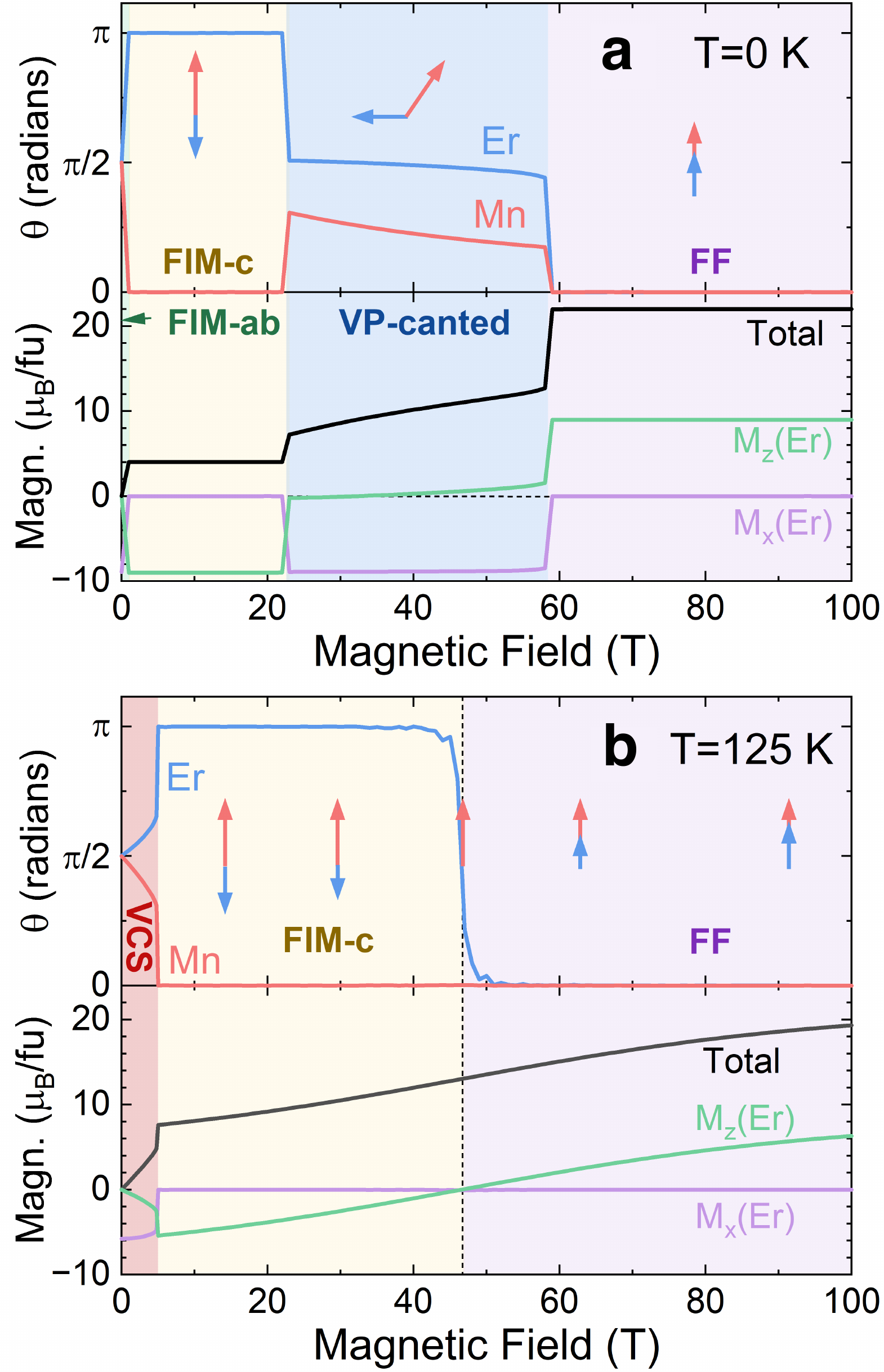}
		\centering
		\caption{Evolution of different magnetic phases obtained from the minimization of the free energy.  Plots show the polar angles and total magnetization as a function of $\mathbf{H}\parallel\mathbf{c}$ for \textbf{a} $T=0$~K and \textbf{b} $T=125$~K. }
		\label{fig:mag_high_SI}
	\end{figure}
	
	{\it Classical energy states with planar anisotropy}.  States with non-zero planar anisotropy or under in-plane magnetic fields can adopt distorted magnetic structures. The moment directions can vary in a complex manner from layer-to-layer within a large magnetic unit cell determined by the overall periodicity. Rather than attempt to evaluate the self-consistent solutions to the free energy, where molecular fields must be evaluated independently for each layer, we rather minimize the total energy of the form
	\begin{align}
		&E = \sum_{j=1}^p \left(3\mathcal{J}_1^{\text{MM}}s^2  \cos\left(\phi_{2j-1}-\phi_{2j}\right)\right. \\ \nonumber
		& +3\mathcal{J}_2^{\text{MM}} s^2  \cos\left(\phi_{2j}-\phi_{2j+1}\right) \\ \nonumber
		&+ 3\mathcal{J}_3^{\text{MM}}s^2  \left[\cos\left(\phi_{2j-1}-\phi_{2j+1}\right)+\cos\left(\phi_{2j}-\phi_{2j+2}\right)\right] \\ \nonumber
		&+ 6\mathcal{J}^{\text{ME}} sS \left[\cos\left(\phi_{2p+j}-\phi_{2j-1}\right)+\cos\left(\phi_{2p+j}-\phi_{2j}\right)\right]  \\ \nonumber
		&+ K_3^{\prime} \cos\left(6\phi_{2p+j}\right) \\ \nonumber
		&- \mu_B\mu_0H_x \left\{3gs\left[\cos\left(\phi_{2j-1}\right)+\cos\left(\phi_{2j}\right)\right]+g_JJ\cos\left(\phi_{2p+j}\right)\right\} \\ \nonumber
		&\left.- \mu_B\mu_0H_y \left\{3gs\left[\sin\left(\phi_{2j-1}\right)+\sin\left(\phi_{2j}\right)\right]+g_JJ\sin\left(\phi_{2p+j}\right)\right\} \vphantom{3\mathcal{J}_1^{\text{MM}}s^2\cos\left(\phi_{2j-1}-\phi_{2j}\right)}\right).
	\end{align}
	Here, $p$ is the number of unit cells containing an Mn-Er-Mn trilayer and $j$ labels a single unit cell.  Mn and Er moments adopt a planar angle $\phi_i$ in each successive layer in the stack with $i=1$ to $2p$ and $i=2p+1$ to $3p$, respectively, with open boundary conditions.  We find the minimum energy using the exchange parameters in Table~1 of the main text and for different values of the in-plane anisotropy parameter $K_3^{\prime}$.  The mean periodicity is shown in Fig.~6b of the main text for a stack consisting of $p=36$ unit cells ($108$ layers).
	
	\bibliography{ErMn6Sn6_Diffraction.bib}

\end{document}